\documentclass[twoside,journal]{IEEEtran}

\usepackage{amsfonts}
\usepackage{amssymb}
\usepackage{amsmath}
\usepackage{mathrsfs}
\usepackage[dvips]{graphicx}
\usepackage{verbatim}
\usepackage{subfigure}
\usepackage{balance}
\usepackage{booktabs}
\usepackage{fancybox}
\usepackage{bm}
\usepackage{extarrows}
\usepackage{algorithm}
\usepackage{algorithmic}
\usepackage{multirow}
\usepackage{array}
\newtheorem{theorem}{Theorem}
\newtheorem{lemma}{Lemma}
\newtheorem{corollary}{Corollary}
\newtheorem{proof}{Proof}

\begin{document}

\title{Safeguarding Millimeter Wave Communications Against Randomly Located Eavesdroppers}

\author{Ying~Ju,
        Hui-Ming~Wang,~\IEEEmembership{Senior Member,~IEEE,} Tong-Xing~Zheng,~\IEEEmembership{Member,~IEEE,}
        Qinye Yin,
        and~Moon~Ho~Lee,~\IEEEmembership{Life~Senior~Member,~IEEE}
\thanks{Y. Ju is with Xi'an Jiaotong University, Xi'an 710049, China, and also with State Radio Monitoring Center, Xi'an 710200, China (e-mail: juyingtju@163.com).}
\thanks{H.-M. Wang, T.-X. Zheng and Q. Yin are with Xi'an Jiaotong University, Xi'an 710049, China (e-mail: xjbswhm@gmail.com; txzheng@stu.xjtu.edu.cn; qyyin@mail.xjtu.edu.cn).}
\thanks{M. H. Lee is with Chonbuk National University, Jeonju 561-756, South Korea (e-mail: moonho@jbnu.ac.kr).}
}



\maketitle
\begin{abstract}
Millimeter wave offers a sensible solution to the capacity crunch faced by 5G wireless communications. This paper comprehensively studies physical layer security in a multi-input single-output (MISO) millimeter wave system where multiple single-antenna eavesdroppers are randomly located. Concerning the specific propagation characteristics of millimeter wave, we investigate two secure transmission schemes, namely maximum ratio transmitting (MRT) beamforming and artificial noise (AN) beamforming. Specifically, we first derive closed-form expressions of the connection probability for both schemes. We then analyze the secrecy outage probability (SOP) in both non-colluding eavesdroppers and colluding eavesdroppers scenarios. Also, we maximize the secrecy throughput under a SOP constraint, and obtain optimal transmission parameters, especially the power allocation between AN and the information signal for AN beamforming. Numerical results are provided to verify our theoretical analysis. We observe that the density of eavesdroppers, the spatially resolvable paths of the destination and eavesdroppers all contribute to the secrecy performance and the parameter design of millimeter wave systems.
\end{abstract}


\IEEEpeerreviewmaketitle

\section{Introduction}\label{s1}
Driven by an increasing number of smart devices and wireless data applications,  an explosive growth of demand for spectrum  in wireless communications appears during the past years. Exploiting millimeter wave becomes a promising approach for providing plentiful spectrum resources to improve the system capacity [\ref{r1}], [\ref{r2}]. Following this trend, the study on millimeter wave communications has attracted great research affords. Millimeter wave channel modeling [\ref{r3}], [\ref{r5}], beamforming schemes [\ref{r6}]-[\ref{r10}] and network performance [\ref{r41}], [\ref{r42}] have been investigated intensively in the past few years.
It becomes a promising candidate for the 5G cellular system.

Given the open feature of the wireless channels, security is a significant concern when designing wireless transmission schemes.  Physical layer security has become a popular way to improve the secrecy performance of wireless communication systems by utilizing wireless channel characteristics [\ref{r11}]-[\ref{r12}]. Thanks to the application of multi-antenna techniques, physical layer security is greatly enhanced in [\ref{r13}], [\ref{r15}]. With multiple antennas, the transmitter can use transmit beamforming either to enhance the legitimate user's channel, i.e., maximum ratio transmitting (MRT) beamforming [\ref{r16}], or to deteriorate eavesdroppers' channels by emitting artificial noise (AN), i.e., AN beamforming [\ref{r19}], [\ref{r21}]. Also, transmit antenna selection technique can be exploited as an effective approach to improve the quality of the legitimate user's channel [\ref{r47}]. When designing secure transmission schemes, reducing secrecy outage probability (SOP) [\ref{r16}], [\ref{r47}] and increasing secrecy throughput [\ref{r21}] are two significant goals.

In wiretap scenarios, eavesdroppers are always passive and their locations are hard to acquire in practice. To model the unknown locations of potential eavesdroppers, stochastic geometry theory has provided a powerful tool recently, with which eavesdroppers' positions can be represented by a spatial distribution such as a Poisson point process (PPP) [\ref{r22}]-[\ref{r241}]. This makes the secure transmission scheme design and secrecy performance evaluation possible in the wireless systems with potentially unknown eavesdroppers.

We should point out that, the wireless channel significantly influences the design and analysis of physical layer security, and the millimeter wave channel is truly different from the traditional microwave channel which has rich scattering. Based on the measurements conducted in New York City, the ray cluster channel model, constituted by several clusters of propagation paths, is built for millimeter wave systems [\ref{r5}]. This model is further adopted in [\ref{r6}]-[\ref{r9}] to design and analyze millimeter wave beamforming schemes. Therefore, the major concern for the implementation of physical layer security in millimeter wave communication systems is the specific propagation characteristics of millimeter wave, which can be described as follows. Firstly, due to the sparse multipaths and scattering of the millimeter wave propagation environment, traditional statistically independent fading distributions are no longer suitable to model the millimeter wave channel. Channels in the millimeter wave band are correlated fading rather than independent and identically distributed (i.i.d.) Rayleigh. Secondly, the small carrier wavelength of millimeter wave enables the realization of large antenna arrays, which can produce extremely high beamforming gain and directionality [\ref{r26}]. This helps to improve the secrecy performance of millimeter wave transmission [\ref{r29}].

Driven by the new propagation features, studies on secure transmissions in millimeter wave systems spring up, both in point-to-point transmissions [\ref{r28}]-[\ref{r39}] and networks [\ref{r36}]-[\ref{r43}]. Specifically, for the point-to-point millimeter wave systems, switched array techniques are utilized in [\ref{r28}], [\ref{r35}], where a subset of transmit antennas are randomly selected to emit signals with every symbol period. This results in a clear constellation in the legitimate user's direction and a high symbol error rate in undesired directions.
This method that needs only a single RF chain is easy to implement in millimeter wave systems. However, the switching speed to be matched at per-symbol rate leads to a huge system overhead, and the antenna sparsity caused by switching makes the secure transmission vulnerable to attacking [\ref{r40}]. Hybrid  beamforming design for millimeter wave systems to resist eavesdropping is studied in [\ref{r44}], [\ref{r45}]. Furthermore, in our previous work [\ref{r46}]-[\ref{r39}], we design beamforming schemes and analyze secrecy performance for the millimeter wave system which contains only one eavesdropper. For the scope of millimeter wave networks, the authors in [\ref{r36}]-[\ref{r43}] analyze the secrecy performance of cellular or Ad hoc networks under the stochastic geometry framework. Both the noise-limited and AN-assisted cellular networks are considered in [\ref{r36}]. The tradeoff between the connection outage probability and secrecy outage probability is investigated for a microwave and millimeter wave hybrid cellular network in [\ref{r37}]. The impact of random blockages and antenna gain on the secrecy performance of Ad hoc networks is analyzed in [\ref{r43}]. These three works focus on the network-wide performance analysis, and the beam pattern is approximated by a sectored antenna model for mathematical tractability.

In all the aforementioned studies, they either do not  consider multipath transmission, or do not investigate the effect of multiple randomly distributed eavesdroppers.
To the best of our knowledge, no previous work has provided secure transmission schemes and comprehensive secrecy performance analysis under a more practical \emph{ray cluster channel model} that characterizes multipath propagation for a millimeter wave system with the stochastic geometry framework. So far, how to safeguard the point-to-point millimeter wave system against randomly located eavesdroppers under a more practical millimeter wave channel model is still unknown, which motivates our work.
\subsection{Our Work and Contributions}
In this paper, we study physical layer security in a multi-input single-output (MISO) millimeter wave system considering multipath propagation under a stochastic geometry framework, where the locations of multiple single-antenna eavesdroppers are modeled as a homogeneous PPP. Connection probability, SOP and secrecy throughput are studied to evaluate the secrecy performance of the transmission schemes. Our contributions are summarized as follows:

1) In the presence of multiple randomly located eavesdroppers, we investigate two transmission schemes, namely MRT beamforming and AN beamforming, under the discrete angular domain channel model which characterized by multiple spatially resolvable paths. We obtain the probability distribution function (PDF) for the number of overlapped common channel paths between the destination and an arbitrary eavesdropper to facilitate the secrecy performance analysis.

2) We derive the closed-form connection probability for both transmission schemes and evaluate the impact of the number of destination's resolvable paths on the connection. Then we obtain the closed-form expressions of SOP for the non-colluding eavesdroppers scenario and the accurate approximation of SOP for the colluding eavesdroppers scenario. In addition, we maximize the secrecy throughput for both schemes, and derive the optimal power allocation between AN and the information signal for the AN scheme. we observe that more power should be allocated to AN in the dense eavesdroppers scenario or in the situation where the number of the destination's resolvable paths or that of the eavesdropper's resolvable paths is large.

3) We reveal that AN beamforming has a better secrecy performance than MRT beamforming when the number of the eavesdropper's resolvable paths is large, the density of eavesdroppers is large or the transmit power is high. Otherwise, MRT beamforming as a simple method shows its superiority. Furthermore, we find that the decrease of the number of the destination's resolvable paths is beneficial for improving the secrecy performance in both beamforming schemes, while the impact of the number of the eavesdropper's resolvable paths on the secrecy throughput are different between two schemes.

\subsection{Organization and Notations}
This paper is organized as follows. In Section \ref{s2}, we build the channel model, analyze spatially resolvable paths, and describe performance metrics. In Section \ref{s3}, we propose two secure transmission schemes against randomly located eavesdroppers. In Sections \ref{s5} and \ref{s4}, we analyze the connection probability, the SOP and the secrecy throughput for both schemes. In Section \ref{s6}, we provide numerical results to verify our theoretical analysis. In Section \ref{s7}, we conclude our paper.

We use the following notations in this paper: bold uppercase (lowercase) letters denote matrices (vectors). $(\cdot)^*$, $(\cdot)^T$, $(\cdot)^H$, $|\cdot|$, $\|\cdot\|$, $\mathbb{P\{\cdot\}}$ and $\mathbb{E}_A\{\cdot\}$ denote conjugate, transpose, conjugate transpose, absolute value, Euclidean norm, probability, and mathematical expectation with respect to A, respectively. $\mathcal{CN}(\mu,\sigma^2)$, $\operatorname{Exp}(\lambda)$ and $\operatorname{Gamma}(N,\lambda)$ denote circularly symmetric complex Gaussian distribution with mean $\mu$ and variance $\sigma^2$, exponential distribution with parameter $\lambda$, and gamma distribution with parameters $N$ and $\lambda$, respectively. $\mathbb{C}^{M\times N}$ denotes the space of all $M\times N$ matrices with complex-valued elements. $\mathbb{Z}^+$ denotes positive integer domain. $\operatorname{log}(\cdot)$, $\operatorname{lg}(\cdot)$ and $\operatorname{ln}(\cdot)$ denote base-2, base-10 and natural logarithms, respectively. $f_u(\cdot)$, $F_u(\cdot)$ and $F_u^{-1}(\cdot)$ denote the PDF, cumulative distribution function (CDF) of $u$ and inverse function of $F_u(\cdot)$, respectively. The intersection, union and difference
between two sets $\Omega_1$ and $\Omega_2$ are denoted by $\Omega_1\cap\Omega_2$, $\Omega_1\cup\Omega_2$ and $\Omega_1 \backslash \Omega_2$, respectively. $\operatorname{E}_i(-x)=\int_x^{\infty}\frac{e^{-t}}{t}dt$ with $x>0$.

\section{System Model}\label{s2}
We consider a millimeter wave system where a transmitter communicates with a destination while randomly located eavesdroppers attempt to intercept the information. The transmitter is equipped with $N_{t}$ antennas, the destination and eavesdroppers are all equipped with single antenna\footnote{This assumption is used for tractability. In practice,  multiple receive antennas are equipped and they will form a receive beam, which is equivalent to a directional single antenna. This will not influence the analysis performed in this paper. Similar assumption has also been adopted by [\ref{r6}], [\ref{r28}], [\ref{r35}], [\ref{r45}]-[\ref{r37}], etc.}. Without loss of generality, we assume the transmitter is located at the origin and the destination is located at coordinate ($r_d$, 0). As shown in Fig. \ref{f1}(a), eavesdroppers are located according to a homogeneous PPP $\Phi_e$ of density $\lambda$ on the 2-D plane with the $k^{th}$ eavesdropper having a distance $r_k$ from the transmitter.

\subsection{Discrete Angular Domain Channel Model}
Due to the sparse characteristics of the millimeter wave propagation environment, millimeter wave channels can be described by a ray cluster based spatial channel model [\ref{r5}]-[\ref{r9}]. The channel is assumed to be a sum of the contributions of $N_{c}$ clusters with $N_{r}$ paths in each cluster and can be formulated as $\mathbf{h}=\sqrt{\frac{\beta}{N_{c}N_{r}}}\sum_{l_c} \sum_{l_r} g_{l_c,l_r}\mathbf{a}(\Theta_{l_c,l_r})^H$, where $\beta$ is the average path loss between the transmitter and the receiver, $g_{l_c,l_r}$ is the complex gain of the $l_r^{th}$ path in the $l_c^{th}$ cluster, $\mathbf{a}(\Theta_{l_c,l_r})$ is the normalized array response at the azimuth angle of departure (AOD) of $\theta_{l_c,l_r}$, and $\Theta_{l_c,l_r}\triangleq\sin(\theta_{l_c,l_r})$. When a uniform linear array (ULA) is adopted, the normalized array response can be described as
$\mathbf{a}(\Theta)=\frac{1}{\sqrt{N_t}}\left[1,e^{-j\frac{2\pi d}{\lambda}\Theta},e^{-j2\frac{2\pi d}{\lambda}\Theta},\cdots,e^{-j(N_t-1)\frac{2\pi d}{\lambda}\Theta}\right]^T$, where $d$ is the antenna spacing, $\lambda$ is the wavelength, and generally $d=\frac{\lambda}{2}$.

\begin{figure}
\centering
\subfigure[Locations of randomly distributed eavesdroppers]{
\label{fig:subfig:a} 
\includegraphics[width=2.6in]{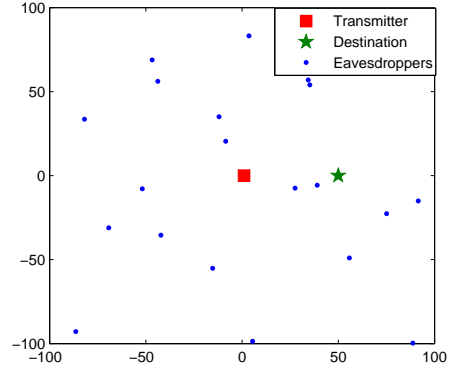}}
\hspace{1in}
\subfigure[Description of spatially resolvable paths]{
\label{fig:subfig:b} 
\includegraphics[width=2.6in]{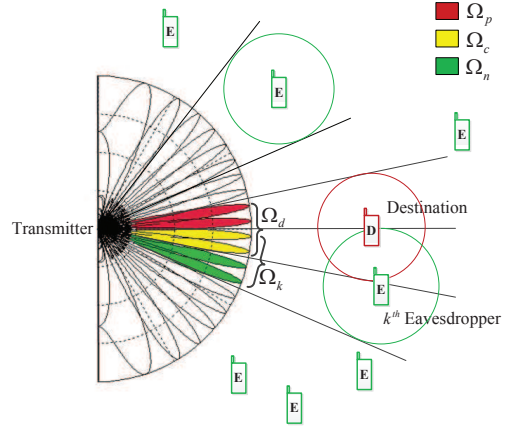}}
\caption{Millimeter wave system model with stochastic geometry framework.} \label{f1}
\label{fig:subfig} 
\end{figure}

Based on the ray cluster model, in order to conduct the theoretical analysis of the transmission schemes, the millimeter wave channel is modeled as a discrete angular domain channel model in existing literature [\ref{r3}], [\ref{r31}], [\ref{r32}] and our previous work [\ref{r46}], which can be described as
\begin{equation}
\mathbf{h}=\sqrt{\frac{1}{L}}r^{-\frac{\alpha}{2}} \mathbf{g} \mathbf{U}^H,
\end{equation}
where $\mathbf{g}=[g_1,g_2,\cdots,g_{Nt}]$ is the complex gain vector, $r^{-\frac{\alpha}{2}}$ is the average path loss, $r$ is the distance between the transmitter and the receiver, $\mathbf{U}\triangleq[\mathbf{a}(\Psi_1),\mathbf{a}(\Psi_2),\cdots,\mathbf{a}(\Psi_{Nt})]$ is the spatially orthogonal basis with $\Psi_{i}\triangleq\frac{1}{M}(i-1-\frac{N_t-1}{2})$ and $M=N_t\frac{d}{\lambda}$. This model is based on the principle that every aperture-limited system has a finite angular resolution [\ref{r31}]. Since paths with $\Theta$ differing by less than $\frac{1}{M}$ are not resolvable by the array, the angular domain can be sampled at a fixed spacing $\frac{1}{M}$ and represented by the spatially orthogonal basis $\mathbf{U}$. Experimental results in [\ref{r5}], [\ref{r30}] show that the millimeter wave channel most likely contains only one cluster where the overwhelming proportion of transmit power is concentrated on. Therefore, we assume that signals are transmitted through one cluster and all the AODs of paths are distributed within the angular range $[\theta_{\operatorname{min}}, \theta_{\operatorname{max}}]$. If $\Psi_i\in \left[ \operatorname{sin}(\theta_{\operatorname{min}}), \operatorname{sin}(\theta_{\operatorname{max}})\right]$, the $i^{th}$ column of $\mathbf{U}$ (the $i^{th}$ orthogonal basis vector) represents a spatially resolvable path and we assume that the $i^{th}$ complex gain $g_i$ is a complex Gaussian coefficient with  $g_i\sim\mathcal{CN}(0,1)$; otherwise, $g_i=0$ [\ref{r5}], [\ref{r26}], [\ref{r46}]. $L$ is defined as the number of spatially resolvable paths with $L<N_t$. Then the channels of the destination and the $k^{th}$ eavesdropper can be described as $\mathbf{h}_d=\sqrt{\frac{1}{L_d}}r_d^{-\frac{\alpha}{2}} \mathbf{g}_d \mathbf{U}^H$ and $\mathbf{h}_k=\sqrt{\frac{1}{L_e}}r_k^{-\frac{\alpha}{2}} \mathbf{g}_k \mathbf{U}^H$, where $L_d$ and $L_e$ are the numbers of the destination's and each eavesdropper's resolvable paths, respectively.

We assume the AODs of all the destination's paths and those of the $k^{th}$ eavesdropper's paths are distributed within the angular range $[\theta_{d,\operatorname{min}},\theta_{d,\operatorname{max}}]$ and $[\theta_{k,\operatorname{min}},\theta_{k,\operatorname{max}}]$ respectively. As shown in Fig. \ref{f1}(b), we define the set $\Omega_d\triangleq \{I_{d,i}|I_{d,i}\in \mathbb{Z}^+,\Psi_{I_{d,i}}\in [\operatorname{sin}(\theta_{d,\operatorname{min}}),\operatorname{sin}(\theta_{d,\operatorname{max}})], I_{d,1}<I_{d,2}<\cdots<I_{d,L_d}\}$, where $I_{d,i}$ is an index of the orthogonal basis vector which represents a destination's spatially resolvable path. Define the set $\Omega_k\triangleq \{I_{k,i}| I_{k,i}\in \mathbb{Z}^+,\Psi_{I_{k,i}}\in [\operatorname{sin}(\theta_{k,\operatorname{min}}),\operatorname{sin}(\theta_{k,\operatorname{max}})], I_{k,1}<I_{k,2}<\cdots<I_{k,L_e}\}$, where $I_{k,i}$ is an index of the orthogonal basis vector which represents a resolvable path of the $k^{th}$ eavesdropper. Define $\Omega=\{1,2,\cdots,N_t\}$, $\Omega_a\triangleq\overline{\Omega}_d=\Omega\backslash\Omega_d$, $\Omega_{c,k}\triangleq\Omega_d\cap\Omega_k$ and $\Omega_{p,k}\triangleq\Omega_d\backslash\Omega_{c,k}$, $\Omega_{n,k}\triangleq\Omega_k\backslash\Omega_{c,k}$. Also, we denote $L_{c,k}$ as the number of the overlapped common paths between the destination and the $k^{th}$ eavesdropper. For notational brevity, we omit $k$ from $L_{c,k}$, $\Omega_{c,k}$, $\Omega_{p,k}$ and $\Omega_{n,k}$, and treat $L_c$, $\Omega_{c}$, $\Omega_{p}$ and $\Omega_{n}$ as functions of $k$ by default. We define the function $\mathcal{S}(\mathbf{B},\Omega_s)$ to generate a matrix whose columns are selected from $\mathbf{B}$, and $\Omega_s$ contains all the selected columns' indexes. Define $\mathbf{g}_{\varrho\nu}\triangleq\mathcal{S}(\mathbf{g}_\varrho,\Omega_\nu)\in\mathbb{C}^{1\times L_\nu}$, where $\varrho\in\{d,k\}$, $\nu\in\{c,p,n,a\}$, and $L_\nu$ is the cardinality of $\Omega_\nu$.

We assume that the instantaneous channel state information (CSI) of the destination is perfectly known at the transmitter [\ref{r19}], [\ref{r21}]. Since eavesdroppers passively receive signals, their instantaneous CSIs are unknown, whereas the distribution of $\mathbf{g}_k$ is available.

\subsection{Spatially Resolvable Paths}
Unlike the traditional wireless channels with rich scattering, the millimeter wave channel involves a limited angular coverage which is represented by the directions of propagation paths. As we have demonstrated in [\ref{r46}], the secrecy performance of the millimeter wave system is dramatically influenced by $L_c$, which is the number of the overlapped common paths between the eavesdropper's and the destination's spatially resolvable paths. When $L_c$ becomes larger, the correlation between the channel of the destination and that of the eavesdropper is larger. More confidential information is leaked to the eavesdropper so that the secrecy performance will be poorer. However, under the stochastic geometry framework, it is hard to get the exact value of $L_c$ due to the randomness of eavesdroppers' locations and the lack of eavesdroppers' CSIs. Fortunately, we derive the PDF of $L_c$ in the following lemma, which will be extensively used in subsequent sections.
\begin{figure}
\centering
\subfigure[$L_c=1$]{
\label{fig:subfig:a} 
\includegraphics[width=2.7in]{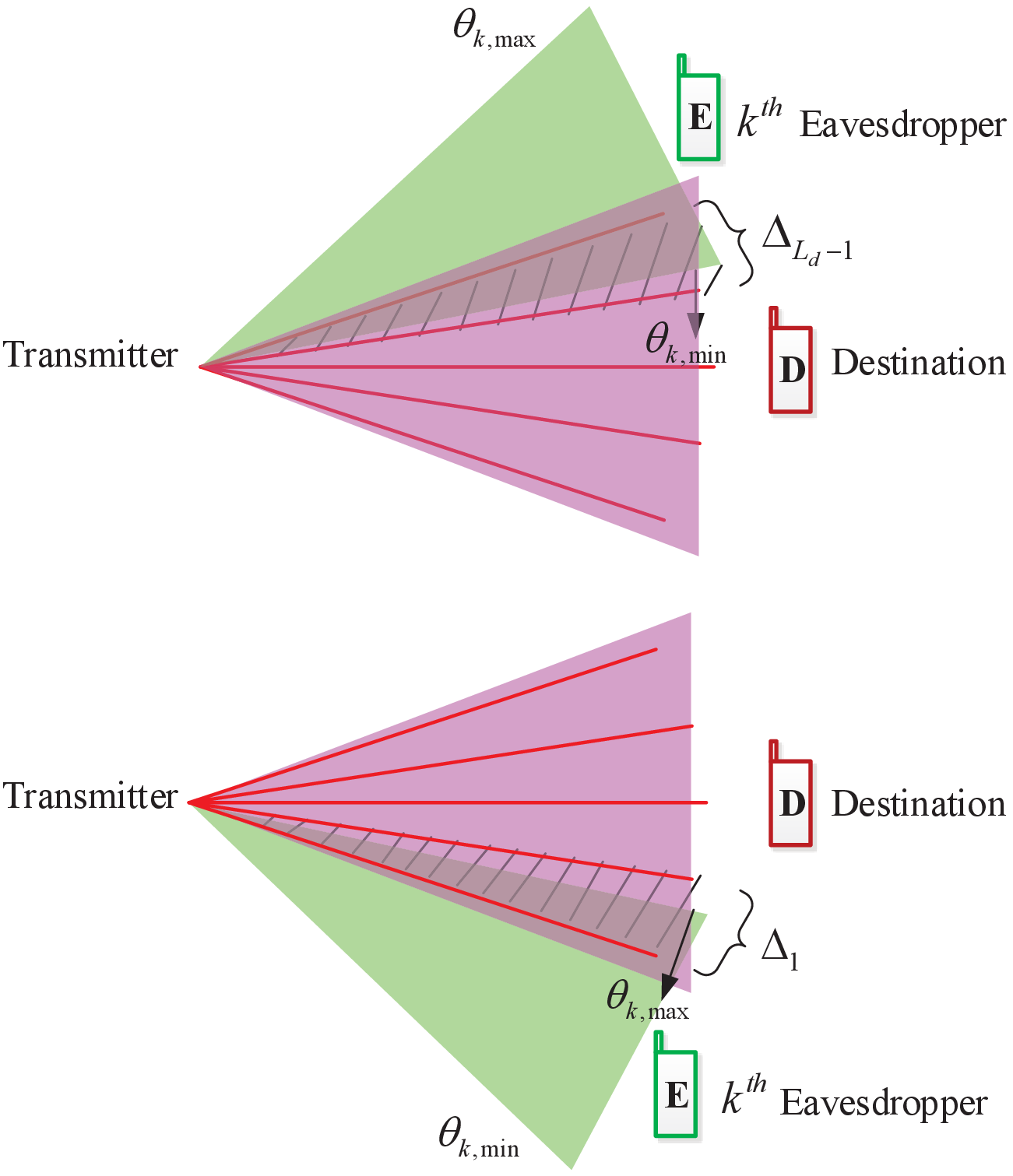}}
\hspace{1in}
\subfigure[$L_c=L_e$($L_d=5,L_e=3$)]{
\label{fig:subfig:b} 
\includegraphics[width=2.7in]{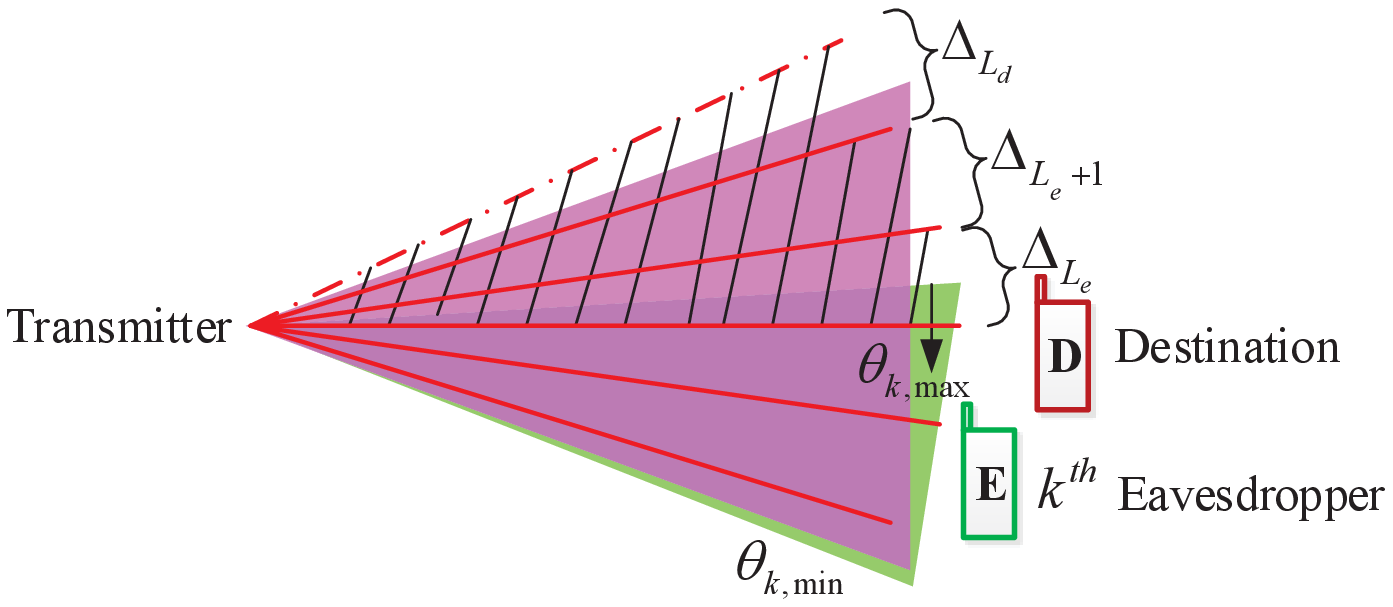}}
\hspace{1in}
\subfigure[$L_c=L_d$($L_d=5,L_e=6$)]{
\label{fig:subfig:b} 
\includegraphics[width=2.7in]{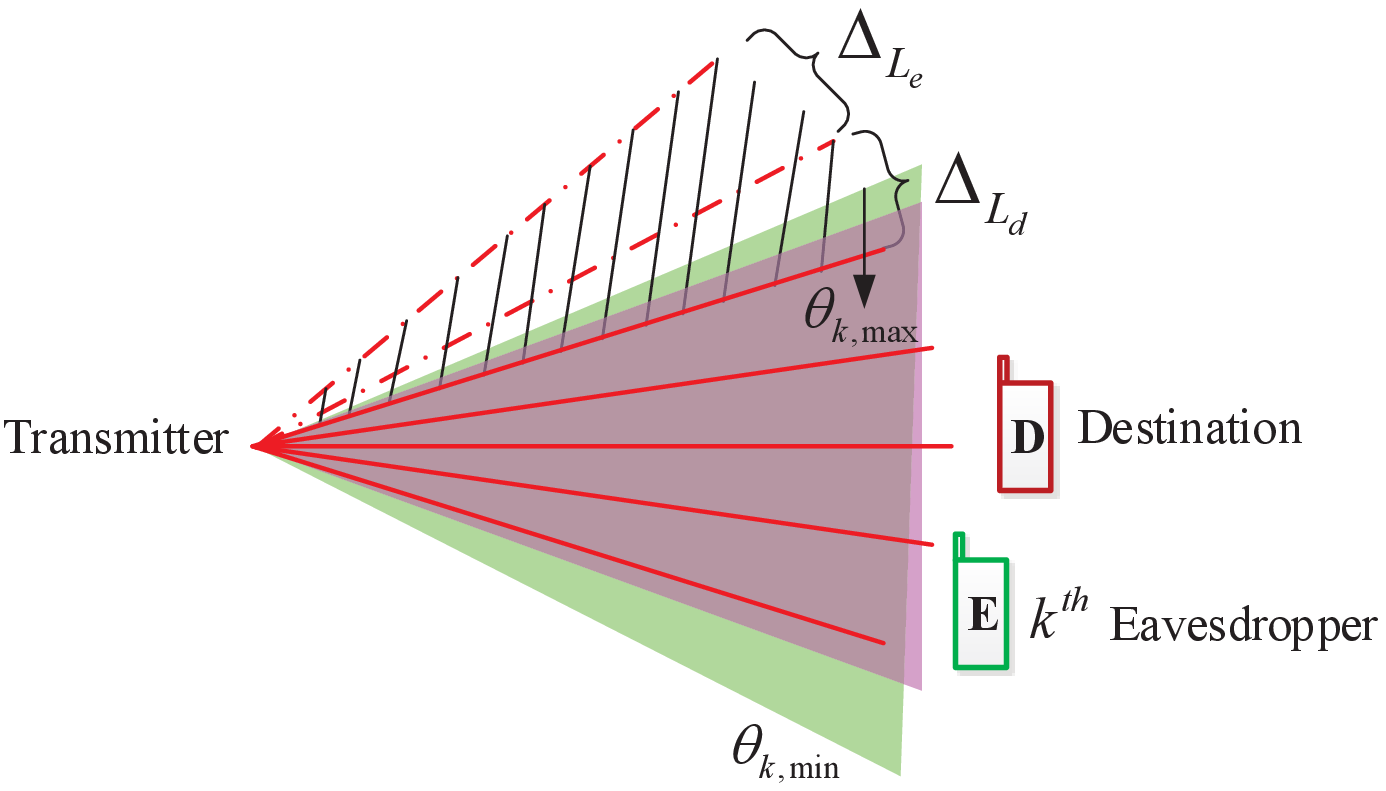}}
\caption{Distribution for the number of the overlapped common paths between the destination and the $k^{th}$ eavesdropper. The pink area shows the angular range $[\theta_{d,\operatorname{min}},\theta_{d,\operatorname{max}}]$ where AODs of the destination's paths are distributed in, and the green area shows the angular range $[\theta_{k,\operatorname{min}},\theta_{k,\operatorname{max}}]$ where AODs of the $k^{th}$ eavesdropper's paths are distributed in. Red solid lines are the spatially resolvable paths of the destination. Each red dashed line represents an angle in $\mathbf{U}$ but not the spatially resolvable path of the destination. (a) describes the situation that $L_c=1$, which is similar as all the situations when $L_c<L_l$. (b) and (c) describe the situation that $L_c=L_l$ with $L_d>L_e$ and $L_d<L_e$ respectively.} \label{f2}
\label{fig:subfig} 
\end{figure}
\lemma{The PDF of $L_c$ can be given by
\begin{equation}
p(L_c)=\left\{ \begin{aligned}
&\frac{\sum_{i=L_l}^{L_u}\omega_{i}}{\pi},\ \ \ \ \ L_c=L_l,\\
&\frac{2\omega_{L_c}}{\pi},\ \ \ \ \ \ \ \ \ \ L_c=1,2,\cdots,L_l-1,\\
&1-\sum_{i=1}^{L_l}p(i), \ \ \ L_c=0,\\
\end{aligned} \right.
\end{equation}
where $\omega_i\triangleq\arcsin\left(\Psi_{\frac{N_t-L_d}{2}+i+1}\right)-\arcsin\left(\Psi_{\frac{N_t-L_d}{2}+i}\right)$, $L_l\triangleq\operatorname{min}\{L_d, L_e\}$ and $L_u\triangleq\operatorname{max}\{L_d, L_e\}$.
}
\begin{proof}
Eavesdroppers are located according to a homogeneous PPP so that the angles of eavesdroppers' locations are uniformly distributed within the range $[-\frac{\pi}{2},\frac{\pi}{2}]$.  In order to get the probability of $L_c=m$, $m\in\{0,1,\cdots,\min\{L_d, L_e\}\}$, we need to know the angular range $\Upsilon_m$ which satisfies that, the destination and the $k^{th}$ eavesdropper will have $m$ overlapped common resolvable paths when the $k^{th}$ eavesdropper is located in $\Upsilon_m$. In other words, $[\theta_{k,\operatorname{min}},\theta_{k,\operatorname{max}}]$ should cover $m$ destination's spatially resolvable paths, where $[\theta_{k,\operatorname{min}},\theta_{k,\operatorname{max}}]$ is the angular range where AODs of the $k^{th}$ eavesdropper' paths are distributed in as discussed in the last subsection. Then we can obtain the probability $p(m)=\frac{\mathcal{W}(\Upsilon_m)}{\pi}$, where $\mathcal{W}(\cdot)$ denotes the width of an angular range. With this idea in mind, we move the $k^{th}$ eavesdropper to find out the angular range $\Upsilon_m$ as shown in Fig. \ref{f2}.

Since the destination is located at ($r_d$, 0) and with $L_d$ resolvable paths, we have $\Omega_d=\noindent\{\frac{N_t-L_d}{2}+1,\frac{N_t-L_d}{2}+2,\cdots,\frac{N_t+L_d}{2}-1,\frac{N_t+L_d}{2}\}$ where $\Omega_{d,i}=\frac{N_t-L_d}{2}+i$. Define the angular range $\Delta_i\triangleq$

\noindent$\left[\operatorname{arcsin}\left(\Psi_{\frac{N_t-L_d}{2}+i}\right),\operatorname{arcsin}\left(\Psi_{\frac{N_t-L_d}{2}+i+1}\right)\right]$,
 and the width $\omega_i\triangleq\mathcal{W}(\Delta_i)=\arcsin\left(\Psi_{\frac{N_t-L_d}{2}+i+1}\right)-\arcsin\left(\Psi_{\frac{N_t-L_d}{2}+i}\right)$. We find that $\Delta_i$ describes the angular range between the $i^{th}$ and the $(i+1)^{th}$ spatially resolvable paths of the destination when $1\leq i\leq L_d-1$. Due to the symmetry of the sine function and the definition of $\Psi_i$, we have $\omega_i=\omega_{L_d-i}$.

As shown in Fig. \ref{f2}(a), if $\theta_{k,\operatorname{max}}$  is located within $\Delta_1$ or $\theta_{k,\operatorname{min}}$ is located within $\Delta_{L_d-1}$, we have $L_c=1$. Thus we derive the PDF of $L_c=1$, which is $p(1)=\frac{\omega_1+\omega_{(L_d-1)}}{\pi}=\frac{2\omega_1}{\pi}$. Then we analyze two different cases $L_d\geq L_e$ and $L_d<L_e$.

1) $L_d\geq L_e$

In this case, $0\leq L_c\leq L_e$. From the analysis of $L_c=1$ given above, we derive the PDFs $p(L_c)=\frac{2\omega_{L_c}}{\pi},L_c=1,2,\cdots,L_e-1$ by analogy. The situation when $L_c=L_e$ is special. As shown in Fig. \ref{f2}(b), when $\theta_{k,\max}\in \Delta_{L_e}\cup \Delta_{L_e+1}\cup\cdots\cup\Delta_{L_d}$, $L_c=L_e$, hence we have $p(L_c)=\frac{\sum_{i=L_e}^{L_d}\omega_{i}}{\pi}$.

2) $L_d<L_e$

In this case, $0\leq L_c\leq L_d$, and we derive the PDFs $p(L_c)=\frac{2\omega_{L_c}}{\pi},L_c=1,2,\cdots,L_d-1$. As shown in Fig. \ref{f2}(c), when  $\theta_{k,\max}\in \Delta_{L_d}\cup \Delta_{L_d+1}\cup\cdots\cup\Delta_{L_e}$, $L_c=L_d$, hence we have $p(L_c)=\frac{\sum_{i=L_d}^{L_e}\omega_{i}}{\pi}$.

Combining the above two cases completes the proof.
\end{proof}

\subsection{Wiretap Encoding Scheme and Performance Metrics}\label{ss23}
We consider both non-colluding eavesdroppers and colluding eavesdroppers scenarios. In non-colluding eavesdroppers scenario, each eavesdropper individually decodes confidential messages and the equivalent signal-to-interference-plus-noise ratio (SINR) of the wiretap channel can be expressed as $\xi_e=\operatorname{max}_{e_k\in\Phi_e}\xi_k$, where $\xi_k$ is the received SINR of the $k^{th}$ eavesdropper. In colluding eavesdropper scenario, eavesdroppers jointly decode confidential messages with maximum ratio combining reception and $\xi_e=\sum_{e_k\in\Phi_e}\xi_k$. Then the capacities of the destination's channel and the wiretap channel are $C_d=\operatorname{log}(1+\xi_d)$ and $C_e=\operatorname{log}(1+\xi_e)$. Adopting the well-known Wyner's wiretap encoding scheme, we denote the codeword rate and secrecy rate as $R_t$ and $R_s$. In addition, we define $R_e\triangleq R_t-R_s$ as the rate redundancy to resist the interception. We analyze the following metrics to evaluate the secrecy performance of transmission schemes.

\emph{Connection probability:} Only if $C_d> R_t$, the destination is able to decode the confidential message correctly. This corresponds to a reliable connection event. We define connection probability as
\begin{equation}\label{e23}
\mathcal{P}_{c}=\mathbb{P}\{C_d>R_t\}.
\end{equation}

\emph{Secrecy outage probability:} If $C_e>R_e$, perfect secrecy is broken and a secrecy outage occurs. We adopt an on-off transmission scheme proposed in [\ref{r21}], where the transmitter decides whether to transmit or not based on the instantaneous CSI of the destination. Throughout the paper, for notational brevity, we define $\mu\triangleq\|\mathbf{g}_d\|^2$ as the overall channel gain of the destination and $\delta$ as the transmission threshold. Since the channel gain $\mu$ varies from time to time, the transmitter emits signals only when $\mu>\delta$; otherwise, the transmission suspends. The SOP is defined as
\begin{equation}
\mathcal{P}_{so}=\mathbb{P}\{C_e>R_t-R_s|\mu\},\ \forall \mu>\delta.  \label{e21}
\end{equation}

\emph{Secrecy throughput:} Secrecy throughput is defined as the effective average transmission rate of the confidential message, which is formulated as
\begin{equation}\label{e22}
\tau=\mathbb{E}_{\mu}\left[R_s(\mu)\right],
\end{equation}
where $R_s(\mu)=0$ for $\mu\leq\delta$.
\section{Transmission Schemes}\label{s3}
In this section, we propose two transmission schemes, namely MRT beamforming and AN beamforming, to resist overhearing of multiple randomly located eavesdroppers.
\subsection{MRT Beamforming}
By exploiting MRT beamforming, the signals received at the destination and the $k^{th}$ eavesdropper are $y_d^{MRT}=\sqrt{P}\mathbf{h}_d\mathbf{w}_1s+n_d$ and $y_k^{MRT}=\sqrt{P}\mathbf{h}_k\mathbf{w}_1s+n_k$, where $\mathbf{w}_1=\mathbf{h}_d^H/\|\mathbf{h}_d\|$ is the beamforming vector, $P$ is the total transmit power, $s$ is the information bearing signal with $\mathbb{E}[|s|^2]=1$, $n_d$ and $n_k$ are i.i.d. additive white Gaussian noise with $n_d\sim\mathcal{CN}(0,\sigma_n^2)$ and $n_k\sim\mathcal{CN}(0,\sigma_n^2)$. We define $\mu_{c,k}\triangleq\|\mathbf{g}_{dc}\|^2$ and $\mu_{p,k}\triangleq\|\mathbf{g}_{dp}\|^2$ as the destination's channel gain of the common paths and the non-common paths with the $k^{th}$ eavesdropper. For notational brevity, we omit $k$ from $\mu_{c,k}$ and $\mu_{p,k}$, and treat them as functions of $k$ by default. We easily find that $\mu=\mu_c+\mu_p$. Then the SNRs of the destination and the $k^{th}$ eavesdropper can be respectively described as
\begin{align}
&\xi_d^{MRT}=\frac{P r_d^{-\alpha}}{\sigma_n^2 L_d}\|\mathbf{g}_d \mathbf{U}^H\|^2=c \mu r_d^{-\alpha},\\
&\xi_k^{MRT}=\frac{P r_k^{-\alpha} |\mathbf{g}_k \mathbf{U}^H \mathbf{U} \mathbf{g}_d^H|^2}{\sigma_n^2 L_e \|\mathbf{g}_d \mathbf{U}^H\|^2}=\frac{a \mu_c r_k^{-\alpha} |\mathbf{g}_{kc} \frac{\mathbf{g}_{dc}^H}{\|\mathbf{g}_{dc}\|}|^2}{\mu}.\label{e30}
\end{align}
where $a\triangleq\frac{P}{L_e \sigma_n^2}$ and  $c\triangleq \frac{P}{L_d \sigma_n^2}$. We find that $g_{k,i}g_{d,i}^*\neq0$ only if $i\in\Omega_d\cap\Omega_k$, i.e., $i\in\Omega_c$. Thus we have $\mathbf{g}_k\mathbf{g}_d^H=\mathbf{g}_{kc}\mathbf{g}_{dc}^H$.

\subsection{AN Beamforming}
Based on the CSI of the destination, we design the AN beamforming matrix as $\mathbf{W}_2=\mathcal{S}(\mathbf{U},\Omega_a)$ to transmit AN to the null space of the destination's channel. Since $\Omega_a=\Omega\backslash\Omega_d$, we have $\mathbf{h}_d\mathbf{W}_2=\mathbf{0}$, hence the destination is not influenced by AN. We observe that by leveraging the specific propagation characteristics of millimeter wave, we form the null space only through selecting some columns from $\mathbf{U}$, which is really simple to operate. Then signals received by the destination and the $k^{th}$ eavesdropper can be described as $y_d^{AN}=\sqrt{\eta P}\mathbf{h}_d\mathbf{w}_1s+\sqrt{\frac{(1-\eta)P}{N_t-L_d}}\mathbf{h}_d\mathbf{W}_2\mathbf{z}+n_d$ and $y_k^{AN}=\sqrt{\eta P}\mathbf{h}_k\mathbf{w}_1s+\sqrt{\frac{(1-\eta)P}{N_t-L_d}}\mathbf{h}_k\mathbf{W}_2\mathbf{z}+n_k$, where $\mathbf{z}\in\mathbb{C}^{(N_t-L_d)\times1}$ is the AN bearing signal with $\mathbb{E}[\mathbf{z}\mathbf{z}^H]=\mathbf{I}_{N_t-L_d}$, $\eta$ is the power allocation ratio of the information signal power to the total transmit power with $0\leq\eta\leq1$. When $\eta=1$, AN beamforming is equavalent  to MRT beamforming where information signal is transmitted with full power. The SINRs of the destination and the $k^{th}$ eavesdropper can be respectively formulated as
\begin{align}
&\xi_d^{AN}=\frac{\eta P r_d^{-\alpha}}{\sigma_n^2 L_d}\|\mathbf{g}_d \mathbf{U}^H\|^2=\eta c \mu r_d^{-\alpha},\\
&\xi_k^{AN}=\frac{\frac{\eta P r_k^{-\alpha}}{L_e\mu} |\mathbf{g}_k \mathbf{U}^H \mathbf{U} \mathbf{g}_d^H|^2}{\frac{(1-\eta) P r_k^{-\alpha}}{(N_t-L_d)L_e} \|\mathbf{g}_k \mathbf{U}^H \mathbf{W}_2 \|^2+\sigma_n^2}\nonumber\\
&\ \ \ \ \ =\frac{\frac{\eta a \mu_c r_k^{-\alpha}}{\mu} |\mathbf{g}_{kc} \frac{\mathbf{g}_{dc}^H}{\|\mathbf{g}_{dc}\|}|^2}{\frac{(1-\eta) a  r_k^{-\alpha}}{(N_t-L_d)} \|\mathbf{g}_{kn}\|^2+1}.\label{e31}
\end{align}
By denoting $\mathbf{g}_k \mathbf{U}^H \mathbf{W}_2=\left[\chi_{1},\chi_{2},\cdots,\chi_{(N_t-L_d)}\right]$, we have  $\chi_{j}=\sum_{i=1}^{N_t}g_{k,i}\mathbf{a}(\Psi_i)^H\mathbf{w}_{2,j}$. Since $\mathbf{W}_2=\mathcal{S}(\mathbf{U},\Omega_a)$ and $\mathbf{U}$ is a unitary matrix, we get $\mathbf{g}_k \mathbf{U}^H \mathbf{W}_2=\mathbf{g}_{ka}$. Since $\Omega_n=\Omega_k\backslash\Omega_c$, by getting rid of those zero elements in $\mathbf{g}_{ka}$, we have $\|\mathbf{g}_{ka}\|=\|\mathbf{g}_{kn} \|$.

\section{Secrecy Performance of MRT Beamforming}\label{s5}
In this section, we analyze the secrecy performance in terms of the connection probability, the SOP and the secrecy throughput for MRT beamforming.
\subsection{Connection Probability}
Since $\mu\sim\operatorname{Gamma}(L_d,1)$, connection probability of MRT beamforming defined in (\ref{e23}) can be described as
\begin{equation}\label{e63}
\begin{aligned}
\mathcal{P}_{c}&=\mathbb{P}\{C_d>R_t\}=\mathbb{P}\{\xi_d^{MRT}>2^{R_t}-1\}\\
&=\frac{1}{\Gamma(L_d)}\Gamma\left(L_d,\frac{(2^{R_t}-1)r_d^\alpha}{c}\right),
\end{aligned}
\end{equation}
where $\Gamma(\cdot)$ is the gamma function, and $\Gamma(\cdot,\cdot)$ denotes the upper incomplete gamma function with $\Gamma(n,x)=(n-1)!\  e^{-x}\sum_{m=0}^{n-1}\frac{x^m}{m!}$.
\subsection{Secrecy Outage Performance}
\subsubsection{Non-Colluding Eavesdroppers}
We first derive the CDF of $\xi_k^{MRT}$ in (\ref{e30}). Define $u\triangleq \left|\mathbf{g}_{kc} \frac{\mathbf{g}_{dc}^H}{\|\mathbf{g}_{dc}\|}\right|^2$, Since each element of $\mathbf{g}_{kc}$ follows a Gaussian distribution with zero mean and unit variance, which is independent of the unit-norm vector $\frac{\mathbf{g}_{dc}^H}{\|\mathbf{g}_{dc}\|}$, we have $u\sim \operatorname{Exp}(1)$. The CDF of $\xi_k^{MRT}$ is given by
\begin{equation}\label{e60}
\begin{aligned}
&F_{\xi_k}(x)=\mathbb{P} \left\{\frac{a \mu_c r_k^{-\alpha}}{\mu} u<x \right\}
=1-e^{-\frac{\mu x r_k^{\alpha}}{a \mu_c}}.
\end{aligned}
\end{equation}

From the definition of $\mu_c$, we find that $\mu_c\sim\operatorname{Gamma}(L_c,1)$. Then the CDF of $\xi_e$ can be calculated as
\begin{align}\nonumber
&F_{\xi_e}(x)=\mathbb{P}\left\{\underset{e_k\in\Phi_e}{\operatorname{max}}\  \xi_k<x \right\}=\mathbb{E}_{\Phi_e,\{\mu_c\}}\left[\underset{e_k\in\Phi_e}{\prod}\mathbb{P}\left\{\xi_k<x\right\}\right]\nonumber\\
&\overset{(a)}{=}\operatorname{exp}\left\{-\lambda\int_0^{\infty}\int_0^{2\pi}\mathbb{E}_{\mu_c}\left\{e^{-\frac{\mu x r_k^{\alpha}}{a \mu_c}}\right\}r_k d\theta dr_k \right\}\nonumber\\
&=\operatorname{exp}\left\{-\pi\lambda\Gamma\left(\frac{2}{\alpha}+1\right)\left(\frac{\mu x}{a}\right)^{-\frac{2}{\alpha}}\sum_{L_c=1}^{L_l} p(L_c)\right.\nonumber\\
&\ \ \ \ \left.\times\mathbb{E}_{\mu_c}\left\{\left(\frac{1}{\mu_c}\right)^{-\frac{2}{\alpha}} \right\} \right\}\nonumber
\end{align}
\begin{align}\label{e61}
&=\operatorname{exp}\left\{-\pi\lambda\Gamma\left(\frac{2}{\alpha}+1\right)\left(\frac{\mu x}{a}\right)^{-\frac{2}{\alpha}}\sum_{L_c=1}^{L_l} p(L_c)\frac{\Gamma(L_c+\frac{2}{\alpha})}{\Gamma(L_c)} \right\},
\end{align}
where $(a)$ holds for the probability generating functional lemma (PGFL) over PPP [\ref{r34}]. We need to mention that when $L_c=0$, $\mathbf{g}_{kc}$ and $\mathbf{g}_{dc}^H$ in (\ref{e30}) both equal to $0$, hence we have $\xi_k=0$ and $\mathbb{P}\{\xi_k<x\}=1$. Since the formula $\underset{e_k\in\Phi_e}{\prod}\mathbb{P}\{\xi_k<x\}$ is a multiplication operation, the case $L_c=0$ does not contribute to $F_{\xi_e}$. Therefore, we only consider $L_c$ from $1$ to $L_l$.

With the CDF of $\xi_e$, we obtain the SOP defined in (\ref{e21}) as
\begin{equation}
\begin{aligned}
&\mathcal{P}_{so}(\mu)= \mathbb{P}\{C_e>C_d-R_s|\mu\}\\
&=\mathbb{P} \left\{\xi_e^{MRT}>\frac{\xi_d^{MRT}-(T-1)}{T}\right\}\\
&=1-\operatorname{exp}\left\{-\pi\lambda\Gamma\left(\frac{2}{\alpha}+1\right)\left(\frac{\mu[ c\mu r_d^{-\alpha}-(T-1)]}{ a T}\right)^{-\frac{2}{\alpha}}\right.\\
&\ \ \ \ \left.\times\sum_{L_c=1}^{L_l} p(L_c)\frac{\Gamma(L_c+\frac{2}{\alpha})}{\Gamma(L_c)} \right\}.
\end{aligned}
\end{equation}
where $T\triangleq 2^{R_s}$.

\subsubsection{Colluding Eavesdroppers}
In this scenario, secrecy outage probability can be written by $\mathcal{P}_{so}(\mu)=\mathbb{P}\left\{\frac{a}{\mu} \sum_{e_k\in\Phi_e}\mu_c u r_k^{-\alpha}>\frac{ c\mu r_d^{-\alpha}-(T-1)}{T}\right\}=\mathbb{P}\left\{I_e>\frac{\mu[ c\mu r_d^{-\alpha}-(T-1)]}{ a T}\right\}$, where $I_e\triangleq\sum_{e_k\in\Phi_e}\mu_c u r_k^{-\alpha}$. We first calculate the Laplace transform of $I_e$ by
\begin{align}\label{e65}
&\mathcal{L}_{I_e}(s)=\mathbb{E}_{\Phi_e,u,\mu_c}\left[ e^{-s\sum_{e_k\in\Phi_e}\mu_c u r_k^{-\alpha}} \right]\nonumber\\
&=\mathbb{E}_{\Phi_e}\left\{ \underset{e_k\in\Phi_e}{\prod} \mathbb{E}_{u,\mu_c}\left[ \exp\left(-s\mu_c u r_k^{-\alpha}\right)\right] \right\}\nonumber\\
&\overset{(b)}{=}\exp\left\{-\lambda\sum_{L_c=1}^{L_l} 2\pi p(L_c)\mathbb{E}_{u,\mu_c}\right.\nonumber\\
&\ \ \ \ \left.\times\left[\int_0^{\infty} \left[1-\exp\left(-s\mu_c u r_k^{-\alpha}\right)\right] r_k dr_k\right]\right\}\nonumber\\
&=\exp\left\{-\pi\lambda\sum_{L_c=1}^{L_l} p(L_c)s^{\frac{2}{\alpha}} \Gamma(1-\frac{2}{\alpha}) \mathbb{E}_{u}\left(u^{\frac{2}{\alpha}}\right)  \mathbb{E}_{\mu_c}\left(\mu_c^{\frac{2}{\alpha}}\right) \right\}\nonumber\\
&=\exp\left\{-\pi\lambda s^{\frac{2}{\alpha}} \Gamma(1-\frac{2}{\alpha}) \Gamma(1+\frac{2}{\alpha})\sum_{L_c=1}^{L_l} p(L_c) \frac{\Gamma(L_c+\frac{2}{\alpha})}{\Gamma(L_c)} \right\},
\end{align}
where $(b)$ holds for PGFL over PPP.

After deriving $\mathcal{L}_{I_e}(s)$, we can obtain the CDF of $I_e$ through inverse Laplace transform, and then get the SOP. However, inverse Laplace transform causes considerable calculation complexity and induces analysis intractable. Therefore, we provide an approximation of $\mathcal{P}_{so}$ in the following theorem.

\theorem{The SOP in the scenario of colluding eavesdroppers for MRT beamforming can be approximated as
\begin{equation}\label{e64}
\mathcal{P}_{so}\gtrapprox \sum_{n=0}^N {N \choose n} (-1)^n \mathcal{L}_{I_e}\left(\frac{ q a T n}{\mu[ c\mu r_d^{-\alpha}-(T-1)]} \right),
\end{equation}
where $\mathcal{L}_{I_e}(s)$ is given by (\ref{e65}), $q\triangleq N(N!)^{-\frac{1}{N}}$ and $N$ is defined as the number of terms used in approximation.
}
\begin{proof}
Defining $\iota$ as a normalized gamma random variable with the shape parameter $N$, we have
\begin{equation}
\begin{aligned}
\mathcal{P}_{so}(\mu)&=\mathbb{P}\left\{\frac{ a T I_e}{\mu[ c\mu r_d^{-\alpha}-(T-1)]}>1\right\}\\
&\overset{(c)}{\approx} \mathbb{P}\left\{\frac{ a T I_e}{\mu[ c\mu r_d^{-\alpha}-(T-1)]}>\iota \right\}\\
&\overset{(d)}{\gtrapprox} \left\{ 1-\exp\left[-\frac{ q a T I_e}{\mu[ c\mu r_d^{-\alpha}-(T-1)]}\right]\right\}^N,
\end{aligned}
\end{equation}
where $(c)$ holds for the fact that a normalized gamma random variable converges to identity when its shape parameter goes to infinity, and $(d)$ follows the CDF bound of a normalized gamma random variable [\ref{r41}], [\ref{r36}]. By using binomial expansion, we obtain the tight lower bound of $\mathcal{P}_{so}$ given in (\ref{e64}).
\end{proof}

\begin{figure}[!t]
\centering
\includegraphics[width=2.9 in]{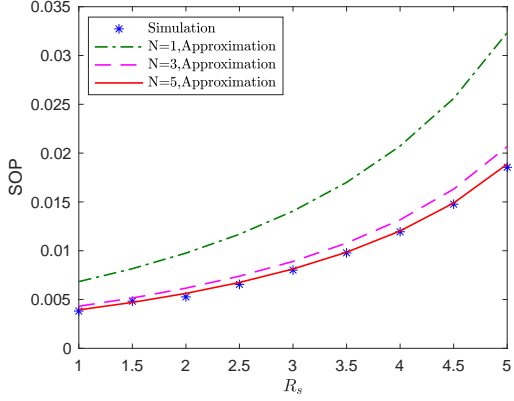}
\caption{SOP versus $R_s$ for different $N$'s, with $N_t=100$, $P=0$dBm, $r_d=50$m, $L_d=20$, $L_e=20$, $\alpha=4$ and $\lambda=10^{-5}$.}\label{f11}
\end{figure}

As shown in Fig. \ref{f11},  $\mathcal{P}_{so}$ given in Theorem 1 is more similar to the simulation results when $N$ becomes larger, and we get a good approximation when $N$ increases to $5$. Thus we use $N=5$ in the following numerical analysis. In addition, we find that  $\mathcal{P}_{so}$ increases when $R_s$ increases. The underlying reason is that when we set a higher $R_s$, the rate redundancy $R_e=R_t-R_s$ which is utilized to against the eavesdropping becomes lower and the secrecy outage requirement $C_e>R_e$ is easier to meet.

\subsection{Secrecy Throughput Maximization}
In this subsection, we maximize the secrecy throughput subject to a tolerable SOP constraint in the non-colluding eavesdroppers scenario. We first maximize $R_s(\mu)$ and formulate the optimization problem as
\begin{subequations}\label{e50}
\begin{align}
&\underset{R_t,\delta}{\operatorname{max}}\ \ R_s(\mu),\nonumber \\
&s.t. \ \ \ 0<R_s(\mu)<R_t\leq C_d,\label{e51} \\
&\ \ \ \ \ \ \ \mathcal{P}_{so}(\mu)\leq \epsilon,\label{e52}
\end{align}
\end{subequations}
where $\epsilon\in[0,1]$ is the SOP threshold. Following the definition formula of the SOP in (\ref{e21}), the SOP constraint can be rewritten as $\mathcal{P}_{so}(\mu)=1-F_{\xi_e}(2^{R_t-R_s}-1)\leq \epsilon$. Since $F_{\xi_e}(x)$ is a monotonically increasing function, we obtain $R_s \leq R_t-\operatorname{log}(1+F_{\xi_e}^{-1}(1-\epsilon))$, where $F_{\xi_e}^{-1}(\cdot)$ denotes the inverse function of $F_{\xi_e}(\cdot)$. As $R_t$ should not exceed $C_d$, we set $R_t$ equal to $C_d$ to achieve a maximum $R_s$, which is
\begin{equation}\label{e62}
R_s^*(\mu)=[C_d-\operatorname{log}(1+F_{\xi_e}^{-1}(1-\epsilon))]^+.
\end{equation}

From (\ref{e61}), by denoting $\rho\triangleq F_{\xi_e}^{-1}(1-\epsilon)$, we have $\rho=\frac{a\varpi}{\mu}$ with $\varpi\triangleq\left[-\frac{\pi\lambda\Gamma\left(\frac{2}{\alpha}+1\right)}{\operatorname{ln}(1-\epsilon)}\sum_{L_c=1}^{L_l} p(L_c)\frac{\Gamma(L_c+\frac{2}{\alpha})}{\Gamma(L_c)}\right]^{\frac{\alpha}{2}}$. In addition, in order to obtain a positive $R_s^*$, $C_d>\operatorname{log}(1+\rho)$ should be satisfied. Since $C_d=\operatorname{log}(1+c \mu r_d^{-\alpha})$, the maximum $R_s(\mu)$ of MRT beamforming can be given by
\begin{equation}
R_s^*(\mu)=\operatorname{log}\frac{1+c \mu r_d^{-\alpha}}{1+\rho},\label{e511}
\end{equation}
with the transmission constraint $\mu>\delta=\sqrt{\frac{z_1 r_d^\alpha}{c}}$, where $z_1\triangleq\varpi a=\rho \mu$, and $\delta$ is the transmission threshold. According to the on-off transmission scheme, the transmitter radiates signals with $R_s^*(\mu)$ only when $\mu>\delta$; otherwise, the transmitter keeps silence and we set $R_s(\mu)=0$. After obtaining $R_s^*(\mu)$, we calculate the maximum secrecy throughput according to (\ref{e22}).

\theorem{The maximum secrecy throughput of MRT beamforming can be given by
\begin{equation}
\begin{aligned}
\tau^*=&\sum_{m=0}^{L_d-1} \frac{e^{-\delta} \delta^{L_d-1-m}}{\Gamma(L_d-m)}  \left[V\left(\frac{1}{\delta}\right)+V\left(\frac{c r_d^{-\alpha}}{1+c r_d^{-\alpha}\delta}\right)\right.\\
&\left.-V\left(\frac{1}{\delta+z_1}\right)+\operatorname{log}\frac{\delta(1+c r_d^{-\alpha}\delta)}{\delta+z_1}\right], \label{e512}
\end{aligned}
\end{equation}
where $V(x)=\frac{1}{\operatorname{ln}2} \sum_{n=1}^{m}\frac{1}{(m-n)!}[\frac{(-1)^{m-n-1}}{x^{m-n}}e^{\frac{1}{x}} \operatorname{Ei}(-\frac{1}{x})+\sum_{k=1}^{m-n}(k-1)!(-\frac{1}{x})^{m-n-k}]$.
}

\begin{proof}
Please see Appendix \ref{a5}.
\end{proof}

\corollary{$\tau^*$ monotonically decreases with $r_d$ and $\lambda$, while monotonically increases with $\epsilon$.}
\begin{proof}
From $\tau^*=\int_\delta^\infty R_s^*(x)f_{\mu}(x)dx=\int_\delta^\infty \operatorname{log}\frac{1+c x r_d^{-\alpha}}{1+\frac{z_1}{x}} f_{\mu}(x)dx$, we have $\frac{\partial \tau^*}{\partial \delta}<0$, $\frac{\partial \tau^*}{\partial z_1}<0$ and $\frac{\partial \tau^*}{\partial r_d}<0$, hence we derive $\frac{d \tau^*}{d z_1}=\frac{\partial \tau^*}{\partial \delta}\frac{d \delta}{d z_1}+\frac{\partial \tau^*}{\partial z_1}<0$. Therefore, we obtain $\frac{d \tau^*}{d \lambda}=\frac{d \tau^*}{d z_1}\frac{d z_1}{d \lambda}<0$, $\frac{d \tau^*}{d \epsilon}=\frac{d \tau^*}{d z_1}\frac{d z_1}{d \epsilon}>0$ and $\frac{d \tau^*}{d r_d}=\frac{\partial \tau^*}{\partial \delta}\frac{d \delta}{d r_d}+\frac{\partial \tau^*}{\partial r_d}<0$.
\end{proof}

Corollary 1 implies that MRT beamforming achieves higher secrecy throughput in a sparser eavesdroppers scenario, under a more moderate SOP constraint, or when the distance between the destination and the transmitter is smaller.

\corollary{At the high transmit power regime, i.e., $P\to \infty$, The maximum secrecy throughput of MRT beamforming is given in (\ref{e5121}), and is independent of $P$.
\begin{equation} \label{e5121}
\tau^*=\sum_{m=0}^{L_d-1} \frac{e^{-\delta} \delta^{L_d-1-m}}{\Gamma(L_d-m)}  V\left(\frac{1}{\delta}\right).
\end{equation}
}
\begin{proof}
When $P\to\infty$, we have $a\to\infty$ and $c\to\infty$. Following (\ref{e511}), $\underset{P\to\infty}{\lim}R_s^*(\mu)=\log\frac{c\mu^2 r_d^{-\alpha}}{a\varpi}=\log\frac{\mu^2 L_e r_d^{-\alpha}}{L_d \varpi}$. By exploiting the same integration method as Theorem 2, we derive the result given in (\ref{e5121}). Since $\delta=\sqrt{\frac{\varpi L_d r_d^{\alpha}}{L_e}}$ is independent of $P$, we have that $\tau^*$ is independent of $P$ when $P\to\infty$.
\end{proof}

\section{Secrecy Performance of AN Beamforming}\label{s4}
In this section, we first investigate the connection and secrecy outage performance of AN beamforming. Then we maximize the secrecy throughput under a given SOP constraint and derive the optimal power allocation ratio $\eta^*$.
\subsection{Connection Probability}
Given $\xi_d^{AN}=\eta c \mu r_d^{-\alpha}$, from (\ref{e63}), the connection probability of AN beamforming can be given by
\begin{equation}
\mathcal{P}_{c}=\frac{1}{\Gamma(L_d)}\Gamma\left(L_d,\frac{(2^{R_t}-1)r_d^\alpha}{\eta c}\right).
\end{equation}
\subsection{Secrecy Outage Performance}
\subsubsection{Non-Colluding Eavesdroppers}
Define $v\triangleq \| \mathbf{g}_{kn} \|^2$, we have $v\sim \operatorname{Gamma}(L_e-L_c,1)$. Since $\Omega_c\cap\Omega_n={\O}$, $u$ and $v$ are independent to each other. The CDF of $\xi_k^{AN}$ can be given by
\begin{equation}\label{e41}
\begin{aligned}
F_{\xi_k}(x)
&=1-\mathbb{E}_v \left[ e^{-\frac{(1-\eta)\mu x}{\eta \mu_c (N_t-L_d)}v-\frac{\mu x r_k^{\alpha}}{\eta a \mu_c}} \right]\\
&\overset{(e)}{=}1-e^{-\frac{\mu x r_k^{\alpha}}{\eta a \mu_c}} \left[1+\frac{(1-\eta)\mu x}{\eta \mu_c (N_t-L_d)}\right]^{-(L_e-L_c)},
\end{aligned}
\end{equation}
where $(e)$ holds for the integration formula [\ref{r33}, 3.326.2]. The CDF of $\xi_e$ can be calculated as
\begin{equation}\label{e47}
\begin{aligned}
&F_{\xi_e}(x)=\mathbb{E}_{\Phi_e,\{\mu_c\}}\left[\underset{e_k\in\Phi_e}{\prod}\mathbb{P}\left\{\xi_k<x\right\}\right]\\
&=\operatorname{exp}\left\{-\lambda\int_0^{\infty}\int_0^{2\pi}\mathbb{E}_{\mu_c}\left\{ \left[1+\frac{(1-\eta)\mu x}{\eta \mu_c (N_t-L_d)}\right]^{-(L_e-L_c)}\right.\right.\\
&\ \ \ \ \left.\left.\times e^{-\frac{\mu x r_k^{\alpha}}{\eta a \mu_c}}\right\}r_k d\theta dr_k \right\}\\
&=\operatorname{exp}\left\{-\pi\lambda\Gamma\left(\frac{2}{\alpha}+1\right)\left(\frac{\mu x}{\eta a}\right)^{-\frac{2}{\alpha}}\sum_{L_c=1}^{L_l} p(L_c)\right.\\
&\ \ \ \ \left.\times \mathbb{E}_{\mu_c}\left\{\left[1+\frac{(1-\eta)\mu x}{\eta \mu_c (N_t-L_d)}\right]^{-(L_e-L_c)} \left(\frac{1}{\mu_c}\right)^{-\frac{2}{\alpha}} \right\} \right\}.
\end{aligned}
\end{equation}

Denoting $b\triangleq \frac{(1-\eta)\mu x}{\eta (N_t-L_d)}$ yields
\begin{equation}\nonumber
\begin{aligned}
&F_{\xi_e}(x)=\operatorname{exp}\left\{-\pi\lambda\Gamma\left(\frac{2}{\alpha}+1\right)\left(\frac{\mu x}{\eta a}\right)^{-\frac{2}{\alpha}}\sum_{L_c=1}^{L_l} p(L_c)\right.\\
&\ \ \ \ \left.\times \int_0^\infty (\mu_c+b)^{-(L_e-L_c)}\mu_c^{L_e-L_c+\frac{2}{\alpha}}\frac{\mu_c^{L_c-1}}{\Gamma(L_c)} e^{-\mu_c} d\mu_c \right\} \\
&\overset{(f)}{=}\operatorname{exp}\left\{-\pi\lambda\Gamma\left(\frac{2}{\alpha}+1\right)\left(\frac{\mu x}{\eta a}\right)^{-\frac{2}{\alpha}}\sum_{L_c=1}^{L_l} \frac{p(L_c)e^b}{\Gamma(L_c)}\right.\\
&\ \ \ \ \left.\times\int_b^\infty y^{-(L_e-L_c)}(y-b)^{L_e+\frac{2}{\alpha}-1} e^{-y} dy \right\}\\
\end{aligned}
\end{equation}
\begin{equation}\label{e48}
\begin{aligned}
&\overset{(g)}{=}\operatorname{exp}\left\{-\pi\lambda\Gamma\left(\frac{2}{\alpha}+1\right)\Gamma\left(\frac{2}{\alpha}+L_e\right)\left(\frac{\mu x}{\eta a}\right)^{-\frac{2}{\alpha}}\right.\\
&\ \ \ \ \left.\times \sum_{L_c=1}^{L_l} \frac{p(L_c)e^{\frac{b}{2}}b^{\frac{L_c+\frac{2}{\alpha}-1}{2}}}{\Gamma(L_c)}W_{\frac{L_c-2L_e-\frac{2}{\alpha}+1}{2},\frac{-L_c-\frac{2}{\alpha}}{2}}(b)\right\},
\end{aligned}
\end{equation}
where $(f)$ follows from the variable transformation $y=\mu_c+b$ and $(g)$ holds for the integration formula [\ref{r33}, 3.383.4]. Since the formula inside the mathematical expectation in (\ref{e47}) is a convex function of $\frac{1}{\mu_c}$, using Jensen's inequality yields
\begin{equation}\label{e49}
\begin{aligned}
&F_{\xi_e}(x)\leq \operatorname{exp}\left\{-\pi\lambda\Gamma\left(\frac{2}{\alpha}+1\right)\left(\frac{\mu x}{\eta a}\right)^{-\frac{2}{\alpha}}\sum_{L_c=1}^{L_l} p(L_c)\right. \\
&\ \left.\times\left[1+\frac{(1-\eta)\mu x}{\eta (N_t-L_d)}\mathbb{E}_{\mu_c}\left(\frac{1}{\mu_c}\right) \right]^{-(L_e-L_c)}
\left[\mathbb{E}_{\mu_c}\left(\frac{1}{\mu_c}\right)\right]^{-\frac{2}{\alpha}} \right\}\\
&=\operatorname{exp}\left\{-\pi\lambda\Gamma\left(\frac{2}{\alpha}+1\right)\left(\frac{\mu x}{\eta a}\right)^{-\frac{2}{\alpha}}\sum_{L_c=1}^{L_l} p(L_c)\left(L_c-1\right)^{\frac{2}{\alpha}}\right.\\
&\ \ \ \ \left.\times \left[1+\frac{(1-\eta)\mu x}{\eta (N_t-L_d)(L_c-1)} \right]^{-(L_e-L_c)}  \right\}.
\end{aligned}
\end{equation}

\begin{figure}[!t]
\centering
\includegraphics[width=2.9 in]{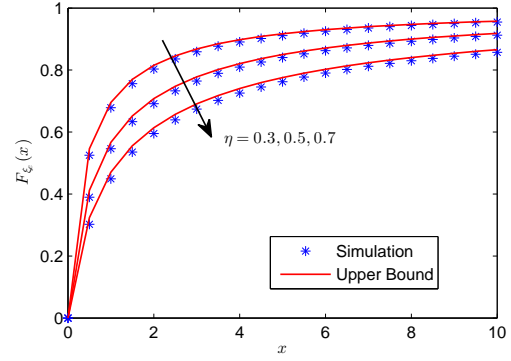}
\caption{$F_{\xi_e}(x)$ versus $x$ for different $\eta$'s, with $N_t=100$,  $L_d=20$, $L_e=20$, $\alpha=4$, $\lambda=1$ and $P=0$dBm.}\label{f3}
\end{figure}

The above formula (\ref{e49}) gives an upper bound of $F_{\xi_e}(x)$. In order to verify the accuracy of (\ref{e49}), we plot $F_{\xi_e}(x)$ in (\ref{e49}) versus $x$ for different values of $\eta$ in Fig. \ref{f3}. We find that the upper-bound curves coincide well with simulation ones, and they are very tight. Thus we use the result in (\ref{e49}) instead of (\ref{e48}) in the following deductions.

Then the SOP of AN beamforming can be expressed as
\begin{equation}\label{e43}
\begin{aligned}
&\mathcal{P}_{so}(\mu)=1-\operatorname{exp}\left\{-\pi\lambda\Gamma\left(\frac{2}{\alpha}+1\right) \right.\\
&\ \ \ \ \times\left\{\frac{\mu[\eta c\mu r_d^{-\alpha}-(T-1)]}{\eta a T}\right\}^{-\frac{2}{\alpha}}\sum_{L_c=1}^{L_l} p(L_c)\left(L_c-1\right)^{\frac{2}{\alpha}}\\
&\ \ \ \ \left.\times\left[1+\frac{\mu(1-\eta) [\eta c \mu r_d^{-\alpha}-(T-1)]}{\eta T(N_t-L_d)(L_c-1)} \right]^{-(L_e-L_c)}  \right\}.
\end{aligned}
\end{equation}

\subsubsection{Colluding Eavesdroppers}
In this scenario, $\mathcal{P}_{so}(\mu)=\mathbb{P}\{\sum_{e_k\in\Phi_e}\frac{\frac{\eta a \mu_c r_k^{-\alpha}}{\mu} u}{\frac{(1-\eta)ar_k^{-\alpha}}{N_t-L_d} v+1}>\frac{ \eta c\mu r_d^{-\alpha}-(T-1)}{T}\}=\mathbb{P}\left\{I_e>\frac{ \eta c\mu r_d^{-\alpha}-(T-1)}{T}\right\}$, where $I_e\triangleq\sum_{e_k\in\Phi_e}\frac{z_2 \mu_c u}{z_3 v+r_k^{\alpha}}$ with $z_2\triangleq\frac{\eta a}{\mu}$ and $z_3\triangleq\frac{(1-\eta)a}{N_t-L_d}$. Then we derive a closed-form expression for $\mathcal{L}_{I_e}(s)$ in the following lemma.
\lemma{
The Laplace transform of $I_e$ can be given by
\begin{equation}\label{e66}
\begin{aligned}
&\mathcal{L}_{I_e}(s)=\exp\left\{- \pi\lambda\Gamma(1+2\alpha)\Gamma(1-2\alpha) \frac{\Gamma(L_e-2\alpha)}{\Gamma(L_e+\frac{3}{2})}z_3^{-2\alpha-1}\right.\\
&\ \ \left.\times z_2s \sum_{L_c=1}^{L_l} p(L_c)L_c F\left(L_c+1,2\alpha+1;L_e+\frac{3}{2};-\frac{sz_2}{z_3}\right) \right\}.\\
\end{aligned}
\end{equation}
}
\begin{proof}
Please see Appendix \ref{a6}.
\end{proof}

Following Theorem 1, in the scenario of colluding eavesdroppers, the SOP of AN beamforming can be approximated as
\begin{equation}
\mathcal{P}_{so}\gtrapprox \sum_{n=0}^N {N \choose n} (-1)^n \mathcal{L}_{I_e}\left(\frac{ q T n}{\eta c\mu r_d^{-\alpha}-(T-1)} \right).
\end{equation}

\subsection{Secrecy Throughput Maximization and Optimal Power Allocation}
In this subsection, we optimize $\eta$ to get a maximum secrecy throughput under the SOP constraint for the AN scheme in the non-colluding eavesdroppers scenario.

According to (\ref{e62}) and by denoting $\rho(\eta)\triangleq\frac{F_{\xi_e}^{-1}(1-\epsilon)}{\eta}$ for the AN scheme, the optimization problem can be formulated as
\begin{equation}\label{e59}
\begin{aligned}
&\underset{\eta}{\operatorname{max}}\ \ R_s(\mu)=\operatorname{log}\frac{1+\eta c \mu r_d^{-\alpha}}{1+\eta \rho(\eta)}, \\
&s.t. \ \ \ 0\leq \eta \leq 1,\ \rho(\eta)<c \mu r_d^{-\alpha}, \\
\end{aligned}
\end{equation}
where $\rho(\eta)<c \mu r_d^{-\alpha}$ is the transmission constraint of the on-off transmission scheme that guarantees a positive $R_s(\mu)$.

Rewriting the definition formula of $\rho(\eta)$ as $F_{\xi_e}^{AN}(\eta \rho(\eta))=1-\epsilon$, we derive $J(\rho)-Q=0$, where $J(\rho)\triangleq\rho^{-\frac{2}{\alpha}}\sum_{L_c=1}^{L_l} z_4[1+z_5(1-\eta)\rho]^{-(L_e-L_c)}$ and $Q\triangleq -\frac{\operatorname{ln}(1-\epsilon)(\frac{\mu}{a})^{\frac{2}{\alpha}}}{\pi\lambda\Gamma(\frac{2}{\alpha}+1)}$, with $z_4\triangleq p(L_c)(L_c-1)^{\frac{2}{\alpha}}$, $z_5\triangleq \frac{\mu}{(N_t-L_d)(L_c-1)}$. It is hard to get a analytical expression of $\rho(\eta)$ for the AN scheme. Instead, we investigate the relationship between $\rho(\eta)$ and $\eta$ in the following lemma in order to find an efficient way to calculate $\rho$ with a given $\eta$.

\lemma{$\rho(\eta)$ is a monotonically increasing and convex function of $\eta$ in the range $\eta\in[0,1]$.}\label{l3}

\begin{proof}
Please see Appendix \ref{a2}.
\end{proof}

From the definition of $\rho$, we easily obtain that $\rho\geq0$. Due to Lamma \ref{l3}, the maximum $\rho$ is achieved at $\eta=1$. Then we can obtain $\rho_{\operatorname{max}}=Q^{-\frac{\alpha}{2}}(\sum_{L_c=1}^{L_l} z_4)^{\frac{\alpha}{2}}$ from $J(\rho)-Q=0$. We define $\Xi(\rho)\triangleq J(\rho)^{-1}-Q^{-1}=0$. Evidently, for an given $\eta$, $\Xi(\rho)$ is a monotonically increasing function of $\rho$. Since $\Xi(0)=-Q^{-1}<0$ and $\Xi(\rho_{\operatorname{max}})=Q^{-1}\{\sum_{L_c=1}^{L_l} z_4\{\sum_{L_c=1}^{L_l} z_4[1+z_5(1-\eta)Q^{-\frac{\alpha}{2}}(\sum_{L_c=1}^{L_l} z_4)^{\frac{\alpha}{2}}]^{-(L_e-L_c)}\}^{-1}-1\}\geq0$, we find that $\Xi(\rho)$ has the unique zero-crossing point. Therefore, we can obtain the unique root $\rho$ of $\Xi(\rho)=0$ by utilizing the bisection method within the range $[0,\rho_{\operatorname{max}}]$.

\theorem{Given $\rho<c \mu r_d^{-\alpha}$, $R_s$ is a concave function of $\eta$. The optimal $\eta^*$ that maximizes $R_s$ is given by
\begin{equation}\label{e510}
\eta^*=\left\{ \begin{aligned}
&1,\ \ \frac{c \mu r_d^{-\alpha}}{1+c \mu r_d^{-\alpha}}-\frac{\rho_{\operatorname{max}}+ \frac{\alpha\mu}{2(N_t-L_d)}\rho_{\operatorname{max}}^2 }{1+\rho_{\operatorname{max}}} >0, \\
&\eta^\star,\ \ \operatorname{otherwise}, \\
\end{aligned} \right. \\
\end{equation}
where $\eta^\star$ is the unique root of the following equation
\begin{equation}\label{e514}
\frac{d R_s}{d\eta}=\frac{1}{\operatorname{ln}2} \left( \frac{c \mu r_d^{-\alpha}}{1+\eta c \mu r_d^{-\alpha}}-\frac{\rho+\eta\frac{d\rho}{d\eta}}{1+\eta \rho} \right)=0,
\end{equation}
with $\frac{d\rho}{d\eta}$ defined in (\ref{ea1}).
}
\begin{proof}
Please see Appendix \ref{a3}.
\end{proof}

The above theorem provides the solution of the optimization problem (\ref{e59}), and supplies an efficient approach, i.e., the bisection method, that can be used to search the optimal $\eta^\star$, due to the concavity of $R_s$ on $\eta$. Then we investigate how the optimal power allocation ratio $\eta^\star$ varies in different scenarios in the following corollary.

\corollary{The optimal $\eta^\star$ monotonically decreases with $\lambda$ and $r_d$, and monotonically increases with $\epsilon$.}
\begin{proof}
Please see Appendix \ref{a4}.
\end{proof}

Corollary 3 indicates that when the transmission is more vulnerable to eavesdropping, i.e., with a poorer quality of the destination's channel, in a denser eavesdroppers scenario or under a more rigorous SOP constraint, we should allocate more power to AN.

Next, we calculate the transmission threshold $\delta$. We have already derived the transmission constraint $\rho(\eta)<c \mu r_d^{-\alpha}$ to guarantee a positive $R_s$. Since $\rho(\eta)$ monotonically increases with $\eta$, if $\rho(0)$, the minimum $\rho(\eta)$, is not below $c \mu r_d^{-\alpha}$, i.e., $\rho(0)\geq c \mu r_d^{-\alpha}\Rightarrow \mu\leq\frac{r_d^{\alpha}\rho(0)}{c}$, the transmission constraint $\rho(\eta)<c \mu r_d^{-\alpha}$ can not be satisfied. Hence we can not find a feasible $\eta$ to maintain the positivity of $R_s$. Therefore, $\mu>\frac{r_d^{\alpha}\rho(0)}{c}$ must be guaranteed, and as a consequence we set $\delta=\frac{r_d^{\alpha}\rho(0)}{c}$, which corresponds to an on-off transmission, i.e., when $\mu>\delta$, the transmitter radiates signals with $R_s^*(\mu)$; otherwise, the transmission suspends and we set $R_s(\mu)=0$.

From (\ref{e22}), the maximum secrecy throughput of AN beamforming can be given by
\begin{equation}\label{e513}
\tau^*=\int_{\frac{r_d^{\alpha}\rho(0)}{c}}^\infty R_s^*(\mu)f_{\mu}(x)dx,
\end{equation}
where $f_{\mu}(x)=\frac{x^{(L_d-1)}}{\Gamma(L_d)}e^{-x}$.

\section{Numerical Results}\label{s6}
In this section, numerical results are presented to verify our theoretical analysis. The transmitter is equipped with an ULA containing 100 antennas. We set the pass loss exponent $\alpha=4$ and the noise power $\sigma_n^2=-60$dBm [\ref{r46}], [\ref{r36}], [\ref{r43}].
\subsection{Connection and Secrecy Outage Performance}
\begin{figure}[!t]
\begin{center}
\includegraphics[width=2.9 in]{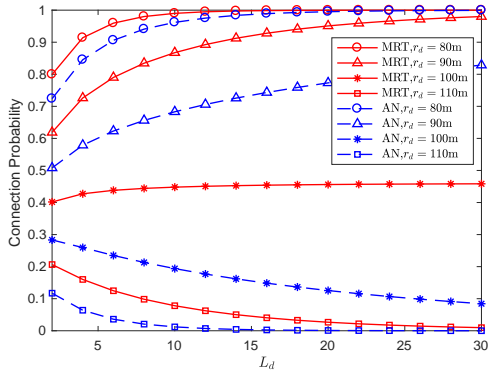}
\end{center}
\caption{Connection probability versus $L_d$ for different $r_d$'s, with $P=10$dBm, $R_t=6$ and $\eta=0.8$.}\label{f12}
\end{figure}

Fig. \ref{f12} describes connection probability versus $L_d$ for different values of $r_d$. We find that when $r_d$ is small, connection probability increases as the number of resolvable paths $L_d$ increases. The trend is opposite when $r_d$ is large. The underlying reason is that when the number of paths becomes larger, the influence of paths with either very high or very low gain becomes weaker, i.e., the received power is closer to its mean and the randomness is smaller. When $r_d$ is small, the mean of SNR is high. Therefore, with the increase of $L_d$, SNR is closer to its higher mean and the connection probability will increase. Otherwise, when $r_d$ is large, SNR approaches to its lower mean and the connection probability will decrease. We also observe that connection probability of AN beamforming is lower than that of MRT beamforming due to the fact that partial transmit power is allocated to the AN transmission.


\begin{figure}[!t]
\begin{center}
\includegraphics[width=2.9 in]{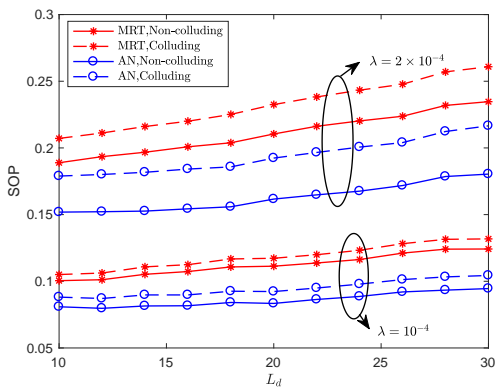}
\end{center}
\caption{SOP versus $L_d$ for different $\lambda$'s, with $N_t=100$, $P=0$dBm, $r_d=50$m, $L_e=20$, $\eta=0.5$ and $R_s=4$.}\label{f4}
\end{figure}

Fig. \ref{f4} presents $\mathcal{P}_{so}$ versus $L_d$ for different values of $\lambda$. As shown in the figure, $\mathcal{P}_{so}$ is higher in colluding eavesdroppers scenario than that in non-colluding eavesdroppers scenario. The differences are more obvious      when eavesdroppers are denser. For both AN beamforming and MRT beamforming, $\mathcal{P}_{so}$ decreases when $L_d$ reduces. The underlying reason is that  when the number of the destination's resolvable paths $L_d$ drops, the information signal is transmitted through fewer directions, which leads to a lower chance of confidential message leakage.

\subsection{Optimal Power Allocation of AN Beamforming}
\begin{figure}[!t]
\begin{center}
\includegraphics[width=2.9 in]{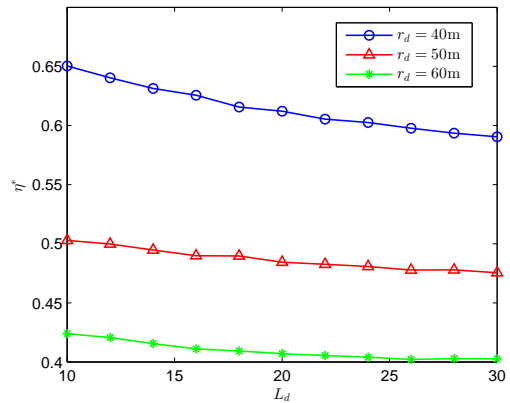}
\end{center}
\caption{Optimal power allocation ratio $\eta^*$ versus $L_d$ for different $r_d$'s, with $N_t=100$,  $P=10$dBm, $L_e=20$, $\lambda=5\times10^{-6}$ and $\epsilon=0.01$.}\label{f5}
\end{figure}

Fig. \ref{f5} investigates the optimal power allocation ratio $\eta^*$ that maximizes the secrecy throughput versus $L_d$ for different values of $r_d$. we find that $\eta^*$ increases with the decrease of $r_d$.
The figure also shows that for a given $L_e=20$, when $L_d$ varies from 10 to 30, $\eta^*$ is getting smaller. It is because that when $L_d$ becomes larger, the transmit beam of the information signals covers more spatially resolvable directions, so that eavesdroppers' resolvable paths are more likely to fall into that beam, which means that eavesdroppers will receive more information signals. Thus we should give a larger fraction of transmit power to AN transmission in order to interfere with eavesdroppers.

\begin{figure}[!t]
\begin{center}
\includegraphics[width=2.9 in]{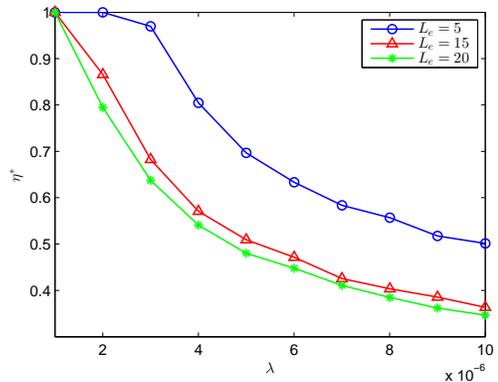}
\end{center}
\caption{Optimal power allocation ratio $\eta^*$ versus $\lambda$ for different $L_e$'s, with $N_t=100$,  $P=10$dBm, $L_d=20$, $r_d=50$m and $\epsilon=0.01$.}\label{f7}
\end{figure}

Fig. \ref{f7} plots the optimal power allocation ratio $\eta^*$ that maximizes the secrecy throughput versus $\lambda$ for different values of $L_e$. As shown in the figure, $\eta^*$ decreases as $\lambda$ increases, which implies that we should increase AN power when secure transmissions become more vulnerable to intercepting. In addition, the value of $\eta^*$ falls down with an increase in $L_e$. The underlying reason is that when $L_e$ becomes larger, resolvable paths of eavesdroppers and those of the destination are more likely to overlap, and then eavesdroppers are able to wiretap more confidential messages. Therefore, more power is ought to be allocated to AN transmission in order to deteriorate eavesdroppers' channels.

\begin{figure}[!t]
\begin{center}
\includegraphics[width=2.9 in]{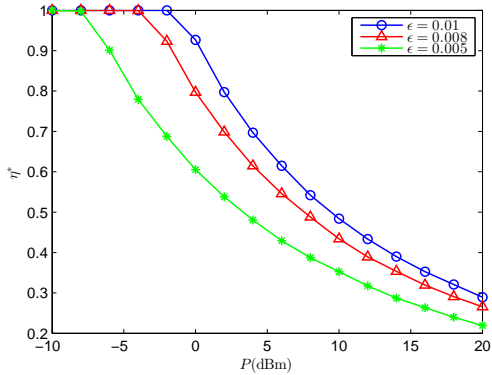}
\end{center}
\caption{Optimal power allocation ratio $\eta^*$ versus $P$ for different $\epsilon$'s, with $N_t=100$,  $P=10$dBm, $L_d=20$, $L_e=20$, $r_d=50$m and $\lambda=5\times10^{-6}$.}\label{f9}
\end{figure}

Fig. \ref{f9} illustrates the optimal power allocation ratio $\eta^*$ that maximizes the secrecy throughput versus $P$ for different values of $\epsilon$. $\eta^*$ keeps 1 at the low transmit power region and then drops as $P$ increases, which implies that we should transmit information signals with full power to achieve a higher message rate at the low power regime and give a larger fraction of power to AN transmission when the transmit power becomes higher. We also find that for a given transmit power $P$, as $\epsilon$ decreases, $\eta^*$ decreases.
\subsection{Secrecy Throughput Performance}
\begin{figure}[!t]
\begin{center}
\includegraphics[width=2.9 in]{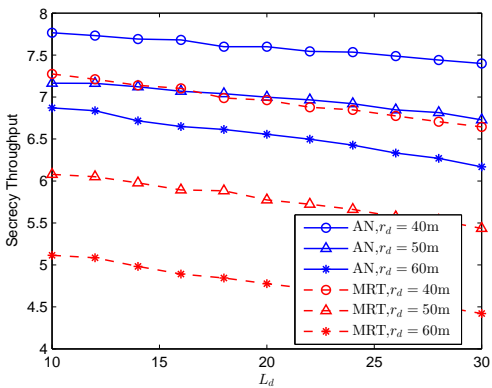}
\end{center}
\caption{Secrecy throughput versus $L_d$ for different $r_d$'s, with $N_t=100$,  $P=10$dBm, $L_e=20$, $\lambda=5\times10^{-6}$ and $\epsilon=0.01$.}\label{f6}
\end{figure}

Fig. \ref{f6} describes the secrecy throughput versus $L_d$ for different values of $r_d$. Secrecy throughput becomes higher with the decrease of $r_d$, which confirms the fact that a better destination's channel contributes to a promotion of secrecy throughput. We also find that the secrecy throughput increases as $L_d$ decreases. The underlying reason is that
confidential signals are radiated through all the directions of the destination's resolvable paths. In the small $L_d$ scenario, the transmit beam which contains secrecy messages is narrow. It is hard for eavesdroppers to intercept. Therefore, smaller $L_d$ is beneficial for enhancing the secrecy throughput.

\begin{figure}[!t]
\begin{center}
\includegraphics[width=2.9 in]{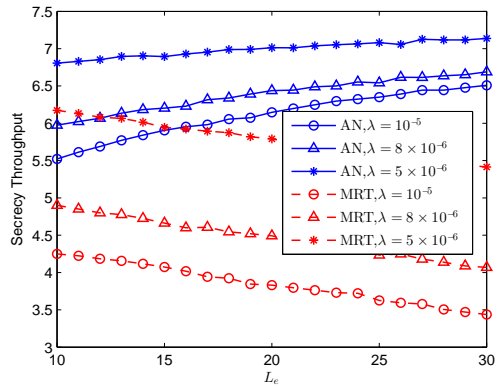}
\end{center}
\caption{Secrecy throughput versus $L_e$ for different $\lambda$'s, with $N_t=100$,  $P=10$dBm, $L_d=20$, $r_d=50$m and $\epsilon=0.01$.}\label{f8}
\end{figure}

Fig. \ref{f8} presents the secrecy throughput versus $L_e$ for different values of $\lambda$. We observe that the secrecy throughput increases with a smaller $\lambda$ for both transmission schemes, which indicates that secrecy performance improves in a sparse eavesdropper scenario. As the number of the eavesdropper's resolvable paths $L_e$ changes, the variation tendencies of MRT beamforming and AN beamforming are different. For MRT beamforming, the secrecy throughput turns to be lower as $L_e$ increases. It is because that in the large $L_e$ situation, eavesdroppers' resolvable paths are more likely to cover the destination's resolvable paths in which directions the information signals emit. Thus eavesdroppers overhear more secrecy messages and the secrecy performance will be poorer. Although a heavier leakage to eavesdroppers also happens in the AN beamforming case when $L_e$ grows, the transmitter gives more power to AN transmission as shown in Fig. \ref{f7}, so that the interference received by eavesdroppers increases too. With the transmission of AN, the SINR of eavesdroppers may drop. Therefore, the results in the figure show that the secrecy throughput increases when $L_e$ increases for the AN scheme. In addition, we observe that the gap between the secrecy throughput of AN beamforming and that of MRT beamforming is more evident in the denser eavesdropper scenario.

\begin{figure}[!t]
\begin{center}
\includegraphics[width=2.9 in]{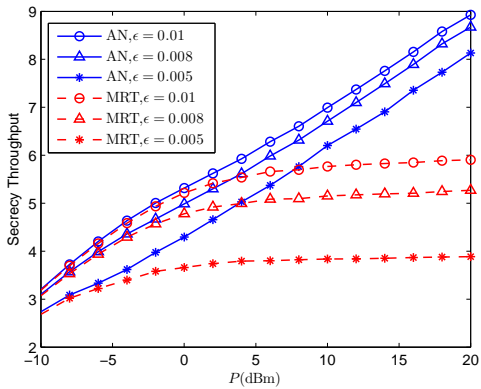}
\end{center}
\caption{Secrecy throughput versus $P$ for different $\epsilon$'s, with $N_t=100$,  $P=10$dBm, $L_d=20$, $L_e=20$, $r_d=50$m and $\lambda=5\times10^{-6}$.}\label{f10}
\end{figure}

Fig. \ref{f10}  describes the secrecy throughput versus $P$ for different values of $\epsilon$. Obviously, by increasing transmit power $P$, the secrecy throughput increases. We also see that MRT beamforming presents a comparable performance to AN beamforming in the low transmit power scenario. As revealed in Corollary 2, the secrecy throughput of MRT beamforming at the high transmit power regime converges to a constant which is irrespective to $P$. However, since more power is allocated to the AN transmission to confuse eavesdroppers as shown in Fig. \ref{f9}, the secrecy throughput of AN beamforming keeps increasing, and the superiority of AN beamforming becomes more obvious. Meanwhile, as the increase of the SOP constraint threshold $\epsilon$, the secrecy throughput becomes higher.
\section{Conclusion}\label{s7}
This paper has comprehensively studied secure transmissions in millimeter wave systems where the locations of eavesdroppers are modeled as an independent homogeneous PPP. We have established a discrete angular domain channel model which characterized by spatially resolvable paths to facilitate the theoretical design and analysis of secure transmission schemes. Then we have evaluated the  performance of MRT beamforming and AN beamforming by obtaining the connection probability, the SOP and the secrecy throughput. Particularly, we have derived the optimal power allocation between AN and the information signal that maximizes the secrecy throughput for AN beamforming. Through our analysis, we have revealed that the superiority of AN beamforming over MRT beamforming is highly significant in dense eavesdroppers scenario, at the high transmit power regime or in the situation that the number of the eavesdropper's spatially resolvable paths is large.

\appendices
\section{Proof of Theorem 2}\label{a5}
Since $\mu\sim\operatorname{Gamma}(L_d,1)$, the PDF of $\mu$ can be given by $f_{\mu}(x)=\frac{x^{(L_d-1)}}{\Gamma(L_d)}e^{-x}$. By substituting (\ref{e511}), we derive the maximum secrecy throughput of MRT beamforming as
\begin{equation}
\begin{aligned}
\tau^* &=\int_\delta^\infty R_s^*(x)f_{\mu}(x)dx\\
&\overset{(h)}{=}\int_0^{\infty}  \frac{1}{\Gamma(L_d)} \operatorname{log}\frac{(y+\delta)[1+c r_d^{-\alpha}(y+\delta)]}{y+\delta+z_1}\\
&\ \ \ \ \times(y+\delta)^{L_d-1}e^{-(y+\delta)}dy\\
&=\frac{e^{-\delta}}{\Gamma(L_d)} \sum_{m=0}^{L_d-1} \binom {L_d-1} m  \delta^{L_d-1-m}\int_0^{\infty} \left[\operatorname{log}(1+\frac{1}{\delta}y)\right.\\
&\ \ \ \ \left.+\operatorname{log}(1+\frac{c r_d^{-\alpha}}{1+c r_d^{-\alpha}\delta}y)-\operatorname{log}(1+\frac{1}{\delta+z_1}y)\right.\\
&\ \ \ \ \left.+\operatorname{log}\frac{\delta(1+c r_d^{-\alpha}\delta)}{\delta+z_1} \right]y^m e^{-y}dy,
\end{aligned}
\end{equation}
where $(h)$ holds for the transformation $y=x-\delta$. According to [\ref{r33}, 4.337.5], the final result shown in (\ref{e512}) is obtained.

\section{Proof of Lemma 2}\label{a6}
\begin{equation}
\begin{aligned}
&\mathcal{L}_{I_e}(s)=\mathbb{E}_{\Phi_e,u,v,\mu_c}\left[ e^{-s\sum_{e_k\in\Phi_e}\frac{z_2 \mu_c u}{z_3 v+r_k^{\alpha}}} \right]\\
&=\exp\left\{- 2\pi\lambda\sum_{L_c=1}^{L_l} p(L_c)\mathbb{E}_{u,v,\mu_c}\left[\int_0^{\infty} \left[1\right.\right.\right.\\
&\ \ \ \ \left.\left.\left. -\exp\left(-s \frac{z_2 \mu_c u}{z_3 v+r_k^{\alpha}} \right) \right]r_k dr_k\right]\right\}\\
&=\exp\left\{- 2\pi\lambda\sum_{L_c=1}^{L_l} p(L_c) \int_0^{\infty} \int_0^{\infty} \int_0^{\infty}s z_2 \mu_c\right.\\
&\ \ \ \ \left.\times(r_k^\alpha+sz_2\mu_c+z_3 v)^{-1}r_k dr_kf(v) f(\mu_c)dvd\mu_c \right\}\\
&\overset{(i)}{=}\exp\left\{- \pi\lambda\Gamma(1+2\alpha)\Gamma(1-2\alpha)\sum_{L_c=1}^{L_l} p(L_c) \right.\\
&\ \ \ \ \left.\times\int_0^{\infty} \int_0^{\infty}  s z_2 \mu_c (sz_2\mu_c+z_3 v)^{-2\alpha-1}f(v) f(\mu_c)dvd\mu_c \right\}\\
&\overset{(j)}{=}\exp\left\{- \pi\lambda\Gamma(1+2\alpha)\Gamma(1-2\alpha) z_3^{-\frac{L_e-L_c}{2}-\alpha}(s z_2)^{\frac{L_e-L_c}{2}-\alpha} \right.\\
&\ \ \ \ \times\sum_{L_c=1}^{L_l} p(L_c)\int_0^{\infty} \mu_c^{\frac{L_e-L_c}{2}-\alpha} e^{\frac{sz_2\mu_c}{2z_3}}\\
&\ \ \ \ \left.\times W_{\frac{-L_e+L_c}{2}-\alpha,\frac{1-L_e+L_c}{2}+\alpha}\left(\frac{sz_2\mu_c}{z_3}\right)f(\mu_c)d\mu_c \right\},\\
\end{aligned}
\end{equation}
where $f(u)=e^{-u}$, $f(v)=\frac{v^{(L_e-L_c-1)}}{\Gamma(L_e-L_c)}e^{-v}$ and $f(\mu_c)=\frac{\mu_c^{(L_c-1)}}{\Gamma(L_c)}e^{-\mu_c}$ are PDFs of $u$, $v$ and $\mu_c$ respectively, $(i)$ holds for the transformation $x=r_k^\alpha$ and the integration formula [\ref{r33}, 3.194], $(j)$ holds for the transformation $y=z_3 v+s z_2\mu_c$ and the integration formula [\ref{r33}, 3.383.4]. By utilizing the integration formula [\ref{r33}, 7.621.3], we obtain the result in (\ref{e66}).

%

\section{Proof of Lemma 3}\label{a2}
The first-order  and second-order derivatives of $\rho(\eta)$ can be given by
\begin{equation}
\frac{d\rho}{d\eta}=-\frac{\partial J/\partial \eta}{\partial J/\partial \rho}
=\frac{\rho^2}{\rho(1-\eta)+\frac{2}{\alpha}\sigma}, \label{ea1}
\end{equation}
\begin{equation}
\begin{aligned}
\frac{d^2 \rho}{d\eta^2}=\frac{2}{\rho}\left(\frac{d\rho}{d\eta}\right)^2
+\frac{\rho^2[\varsigma-\frac{2}{\alpha}\frac{d\sigma}{d\eta}]}{[\rho(1-\eta)+\frac{2}{\alpha}\sigma]^2}, \label{ea2}
\end{aligned}
\end{equation}
where $\sigma\triangleq \frac{\sum_{L_c=1}^{L_l} z_4[1+z_5(1-\eta)\rho]^{-(L_e-L_c)}}{\sum_{L_c=1}^{L_l} z_4 z_5(L_e-L_c)[1+z_5(1-\eta)\rho]^{-(L_e-L_c)-1}}$, and $\varsigma\triangleq \rho-(1-\eta)\frac{d\rho}{d\eta}$. Since $\sigma>0$, obviously, $\frac{d\rho}{d\eta}>0$. By substituting (\ref{ea1}), we have $\varsigma=\frac{\frac{2}{\alpha}\sigma\rho}{\rho(1-\eta)+\frac{2}{\alpha}\sigma}>0$. Denoting $C_1\triangleq \sum_{L_c=1}^{L_l} z_4 z_5(L_e-L_c)[1+z_5(1-\eta)\rho]^{-(L_e-L_c)-1}$ and $C_2\triangleq \sum_{L_c=1}^{L_l} z_4[1+z_5(1-\eta)\rho]^{-(L_e-L_c)}$, we have
\begin{equation}
\begin{aligned}
&\frac{d\sigma}{d\eta}=\frac{\frac{dC_2}{d\eta}}{C_1}-\frac{C_2\frac{dC_1}{d\eta}}{C_1^2}\\
&=\frac{1}{C_1^2}\{C_1\sum_{L_c=1}^{L_l}z_4 z_5(L_e-L_c)[1+z_5(1-\eta)\rho]^{-(L_e-L_c)-1}\varsigma\\
&\ \ \ \ -C_2\sum_{L_c=1}^{L_l}z_4 z_5^2(L_e-L_c)(L_e-L_c+1)\\
&\ \ \ \ \times[1+z_5(1-\eta)\rho]^{-(L_e-L_c)-2}\varsigma\}\\
&<\frac{\varsigma}{C_1^2}\{C_1^2-C_2C_3\}\overset{(k)}{\leq}0,
\end{aligned}
\end{equation}
where $C_3\triangleq \sum_{L_c=1}^{L_l}z_4 z_5^2(L_e-L_c)^2[1+z_5(1-\eta)\rho]^{-(L_e-L_c)-2}$ and $(k)$ holds for the Holder's inequality $(\sum x y)^2\leq (\sum x^2)(\sum y^2)$ with $x=\{z_4[1+z_5(1-\eta)\rho]^{-(L_e-L_c)}\}^{\frac{1}{2}}$ and $y=\{z_4 z_5^2(L_e-L_c)^2[1+z_5(1-\eta)\rho]^{-(L_e-L_c)-2}\}^{\frac{1}{2}}$.
Thus we have that $\frac{d^2 \rho}{d\eta^2}>\frac{2}{\rho}\left(\frac{d\rho}{d\eta}\right)^2>0$.
With $\frac{d \rho}{d\eta}>0$ and $\frac{d^2 \rho}{d\eta^2}>0$, we complete the proof.

\section{Proof of Theorem 3}\label{a3}
The second-order derivative of $R_s$ can be described as
\begin{equation}
\begin{aligned}
&\frac{d^2 R_s}{d\eta^2}
=-\frac{1}{\operatorname{ln}2} \left\{\frac{(c \mu r_d^{-\alpha})^2}{(1+\eta c \mu r_d^{-\alpha})^2}+\frac{1}{(1+\eta \rho)^2}\right.\\
&\ \ \ \ \left.\times\left[(1+\eta\rho)\left(2\frac{d\rho}{d\eta}+\eta\frac{d^2\rho}{d\eta^2}\right)-\left(\rho+\eta\frac{d\rho}{d\eta}\right)^2  \right] \right\},
\end{aligned}
\end{equation}
with $\frac{d\rho}{d\eta}$ and $\frac{d^2\rho}{d\eta^2}$ given by (\ref{ea1}) and (\ref{ea2}). Substituting $\frac{d^2 \rho}{d\eta^2}>\frac{2}{\rho}\left(\frac{d\rho}{d\eta}\right)^2>0$ (see Appendix \ref{a2}) into the above equation yields $\frac{d^2 R_s}{d\eta^2}<-\frac{1}{\operatorname{ln}2}\left[\frac{(c \mu r_d^{-\alpha})^2}{(1+\eta c \mu r_d^{-\alpha})^2}-\frac{\rho^2}{(1+\eta \rho)^2}\right]$. Since $c \mu r_d^{-\alpha}>\rho$, we have $\frac{d^2 R_s}{d\eta^2}<0$, i.e., $R_s$ is a concave function of $\eta$.

For the concavity of $R_s$, the maximum value of $R_s$ is achieved either at the boundaries or at the zero-crossing point of $\frac{d R_s}{d\eta}$. From the first-order derivative formula of $R_s$ in (\ref{e514}), we obtain
$\frac{d R_s}{d\eta}|_{\eta=0}=\frac{1}{\operatorname{ln}2}(c \mu r_d^{-\alpha}-\rho)>0$ and $\frac{d R_s}{d\eta}|_{\eta=1}=\frac{1}{\operatorname{ln}2}\left\{\frac{c \mu r_d^{-\alpha}}{1+c \mu r_d^{-\alpha}}-\frac{\rho_{\operatorname{max}}+ \frac{\alpha\mu}{2(N_t-L_d)}\rho_{\operatorname{max}}^2 }{1+\rho_{\operatorname{max}}}  \right\}$. If $\frac{d R_s}{d\eta}|_{\eta=1}>0$, $R_s$ monotonically increases with $\eta$, and the optimal value of $\eta$ is 1 with the condition directly obtained from  $\frac{d R_s}{d\eta}|_{\eta=1}>0$. If $\frac{d R_s}{d\eta}|_{\eta=1}\leq 0$, $R_s$ first increases and then decreases, and the optimal value of $\eta$ is the unique root of $\frac{d R_s}{d\eta}$.

\section{Proof of Corollary 3}\label{a4}
Substituting (\ref{ea1}) into (\ref{e514}) and denoting $\hat{c}\triangleq c \mu r_d^{-\alpha}$ yield
\begin{equation} \label{ea4}
(\hat{c} \eta^2+1)\rho^2+(\hat{c}\eta+\frac{2}{\alpha}\sigma-\hat{c})\rho-\frac{2}{\alpha}\sigma\hat{c}=0.
\end{equation}
Denote the left side of the above equation as $Y$, we obtain
\begin{equation}
\frac{\partial Y}{\partial\eta}=z_6\frac{d\rho}{d\eta}+ \frac{2}{\alpha}(\rho-\hat{c}) \frac{d\sigma}{d\eta}+2 \hat{c}\eta\rho^2+\hat{c}\rho>0,
\end{equation}
where $z_6\triangleq 2\hat{c}\eta^2\rho+2\rho+\hat{c}\eta+\frac{2}{\alpha}\sigma-\hat{c}$. By substituting (\ref{ea4}), we have $z_6=\rho+\hat{c}\eta^2\rho+\frac{2}{\alpha}\hat{c}\sigma\frac{1}{\rho}>0$. From $\rho<\hat{c}$ and $\frac{d\sigma}{d\eta}<0$, we derive $\frac{\partial Y}{\partial\eta}>0$.

1) $\lambda$: $\frac{d\eta}{d \lambda}=-\frac{\partial Y/\partial\lambda}{\partial Y/\partial \eta}=-\frac{z_6\frac{d\rho}{d\lambda}}{\partial Y/\partial \eta}$, where $z_6>0$ and $\frac{\partial Y}{\partial \eta}>0$. From the definition of $\Xi$, we find $\Xi=0$, $\frac{\partial \Xi}{\partial \rho}>0$ and $\frac{\partial \Xi}{\partial \lambda}<0$, hence $\frac{d\rho}{d\lambda}=-\frac{\partial \Xi/\partial\lambda}{\partial \Xi/\partial \rho}>0$. Thus we derive $\frac{d\eta}{d \lambda}<0$.

2) $\epsilon$: $\frac{d\eta}{d \epsilon}=-\frac{\partial Y/\partial\epsilon}{\partial Y/\partial \eta}=-\frac{z_6\frac{d\rho}{d\epsilon}}{\partial Y/\partial \eta}$, where $z_6>0$ and $\frac{\partial Y}{\partial \eta}>0$. Since $\frac{\partial \Xi}{\partial \epsilon}>0$ , we have $\frac{d\rho}{d\epsilon}=-\frac{\partial \Xi/\partial\epsilon}{\partial \Xi/\partial \rho}<0$. Thus we obtain $\frac{d\eta}{d \epsilon}>0$.

3) $r_d$: $\frac{d\eta}{d r_d}=-\frac{\partial Y/\partial r_d}{\partial Y/\partial \eta}=-\frac{\frac{\partial Y}{\partial \hat{c}}\frac{d\hat{c}}{d r_d}}{\partial Y/\partial \eta}$, where $\frac{d\hat{c}}{d r_d}<0$ and $\frac{\partial Y}{\partial \eta}>0$. Since $\frac{\partial Y}{\partial \hat{c}}=\eta^2\rho^2+\eta\rho-\rho-\frac{2}{\alpha}\sigma$, substituting (\ref{ea4}) yields $\frac{\partial Y}{\partial \hat{c}}=-(\rho^2+\frac{2}{\alpha}\sigma\rho)\frac{1}{\hat{c}}<0$. Thus we obtain $\frac{d\eta}{d r_d}<0$.

With 1)-3), we complete the proof.

%
%
%
%
%
%
%

%
%
%

\end{document}